\documentclass[utf8]{frontiersFPHY}
\setcitestyle{square}

\usepackage[utf8]{inputenc}
\usepackage[english]{babel}
\usepackage{amsmath}
\usepackage{amsthm}
\usepackage{amssymb}
\usepackage{epstopdf}
\usepackage{xcolor,colortbl}
\usepackage{graphicx}
\usepackage{subcaption}
\usepackage{mathrsfs}
\usepackage{textcomp}
\usepackage{placeins}
\usepackage[pdftex, bookmarks, colorlinks, breaklinks]{hyperref}
\usepackage{nameref}
\hypersetup{colorlinks, citecolor=blue}
\usepackage{paralist}

\usepackage{url,hyperref,lineno,microtype,subcaption}
\usepackage[onehalfspacing]{setspace}

\newcommand{\inner}[2]{\langle #1, #2 \rangle}

\newcommand{\R}{\mathbb{R}}

\newcommand{\dx}{\,\mathrm{d}x}
\newcommand{\ds}{\,\mathrm{d}s}

\newcommand{\Jvec}{\textbf{J}}
\newcommand{\nvec}{\textbf{n}}
\newcommand{\norm}[1]{\left\lVert#1\right\rVert}
\newcommand{\Clion}{Cl\textsuperscript{\textminus}}

\newcommand{\Naion}{Na\textsuperscript{+}}
\newcommand{\Kion}{K\textsuperscript{+}}

\newcommand{\Ich}{I_\text{ch}}
\newcommand{\Icap}{I_\text{cap}}
\newcommand{\Icapr}{I_{\text{cap},r}}
\newcommand{\Isyn}{I_\text{syn}}
\newcommand{\Na}{\text{Na}}
\newcommand{\K}{\text{K}}
\newcommand{\Cl}{\text{Cl}}

\DeclareMathOperator{\Div}{\nabla \cdot}
\DeclareMathOperator{\Grad}{\nabla}

\let\emptyset\varnothing

\extraAuth{}
\def\keyFont{\fontsize{8}{11}\helveticabold}
\def\firstAuthorLast{Ellingsrud {et~al.}} 
\def\Authors{A. J. Ellingsrud$^{\dagger, 1}$, A. Solbrå$^{\dagger, 2,3}$, G. T. Einevoll$^{2,3,4}$, G. Halnes$^{\ast, 2, 4}$ and M. E. Rognes$^{\ast, 1}$}
\def\Address{$^{1}$ Department for Scientific Computing and Numerical Analysis, Simula Research Laboratory, Norway \\
$^{2}$Centre for Integrative Neuroplasticity, University of Oslo, Oslo, Norway \\
$^{3}$Department of Physics, University of Oslo, Oslo, Norway \\
$^{4}$Faculty of Science and Technology, Norwegian University of Life Sciences, {\AA}s, Norway\\
  $^{\dagger}$ The first two authors contributed equally. \\
  $^{\ast}$ The last two authors contributed equally. 
}
\def\corrAuthor{M. E. Rognes}
\def\corrEmail{meg@simula.no}

\begin{document}

\onecolumn
\firstpage{1}

\title[Finite element simulation of ionic electrodiffusion]{Finite element simulation of ionic electrodiffusion in cellular geometries}
\author[\firstAuthorLast]{\Authors} 
\address{} 
\correspondance{} 

\maketitle

\begin{abstract}
  Mathematical models for excitable cells are commonly based on cable
  theory, which considers a homogenized domain and spatially constant
  ionic concentrations. Although such models provide valuable insight,
  the effect of altered ion concentrations or detailed cell morphology
  on the electrical potentials cannot be captured. In this paper, we
  discuss an alternative approach to detailed modelling of
  electrodiffusion in neural tissue. The mathematical model describes
  the distribution and evolution of ion concentrations in a
  geometrically-explicit representation of the intra- and
  extracellular domains. As a combination of the electroneutral
  Kirchhoff-Nernst-Planck (KNP) model and the
  Extracellular-Membrane-Intracellular (EMI) framework, we refer to
  this model as the KNP-EMI model. Here, we introduce and numerically
  evaluate a new, finite element-based numerical scheme for the
  KNP-EMI model, capable of efficiently and flexibly handling
  geometries of arbitrary dimension and arbitrary polynomial
  degree. Moreover, we compare the electrical potentials predicted by
  the KNP-EMI and EMI models. Finally, we study ephaptic coupling
  induced in an unmyelinated axon bundle and demonstrate how the
  KNP-EMI framework can give new insights in this setting.

\tiny
 \keyFont{ \section{Keywords:} finite element, electrodiffusion, ion concentrations, cell membrane, ephaptic coupling, KNP-EMI} 

\end{abstract}

\section{Introduction}

The most common computational models for excitable cells are those
based on cable theory~\cite{Rall1977, Koch1999}. In its standard form,
the cable model is based on several simplifying assumptions, most
importantly that the extracellular potential and both intracellular
and extracellular ion concentrations are constant in space and
time. Multi-compartmental neuron models based on cable theory are
widely used within the field of neuroscience to simulate large network
of interacting neurons (see e.g.~\cite{markram2015}). In such models,
only synaptic interactions between neurons are considered, whereas
changes in the extracellular field and extracellular ion
concentrations associated with a neuron's activity are assumed to be
too small to have any influence on its neighboring neurons (or
itself). Although these assumptions are only approximations, the
resulting models still give accurate predictions of neuronal
electrodynamics in many scenarios. Indeed, concentration changes are
often limited by neuronal and glial uptake mechanisms that strive
towards maintaining concentrations close to basal levels.

However, there are also many scenarios that involve dramatic changes
in extracellular ion concentrations. On a large spatial scale, ion
concentration changes are a trademark of several pathological
conditions such as spreading depression or epilepsy~\cite{Dietzel1989,
Somjen2001, Sykova2008, Ayata2015}. Extracellular concentration shifts
will lead to changes in neuronal reversal potentials, and can thus
affect the dynamical properties of the neurons~\cite{Kager2000,
oyehaug_dependence_2011, Wei2014}. Under non-pathological conditions,
concentration-dependent, electrodiffusive effects are hypothesized to
be important in specific microdomains of the
brain~\cite{savtchenko2017electrodiffusion}. In general, the
extracellular ion concentration changes resulting from a neuronal
event can be expected to be largest in regions where the extracellular
space is small and confined.

Similarly, there are several scenarios where the assumption of a
constant extracellular potential may be questionable. For instance,
ephaptic interactions have been reported to play a role for neural
phenomena taking place at both small and large spatial
scales~\cite{Holt1999, bokil2001ephaptic, anastassiou2011ephaptic,
anastassiou2015ephaptic, Goldwyn2016, tveito2017evaluation,
han2018ephaptic, Schifman2019}. Ephaptic interaction (or coupling) is
a coupling between neurons via the extracellular potential, which is
hard or impossible to represent under the aforementioned assumption.

The olfactory nerve is one example in which variations in ion
concentrations and extracellular potentials may be important. Whereas
most axons in the mammalian brain are coated in an insulating layer
of \emph{myelin}, the axons in the olfactory nerve are unmyelinated
and organized in tight
bundles~\cite{doucette1984glial,griff2000ultrastructural}. In view of
the tight packing, one might expect large ion concentration variations
in the extracellular space between the olfactory nerve
axons. Moreover, the olfactory nerve axon arrangement will maximize
any ephaptic coupling, with a potential evolutionary
purpose~\cite{lowe2003electrical}. In addition, diffusion along
extracellular ion concentration gradients can generate so-called
\emph{diffusion potentials}~\cite{halnes2016, savtchenko2017electrodiffusion,
solbra2018kirchoff}, which may constitute an additional ephaptic effect on
membrane potentials.

There are several computational studies considering ephaptic
interaction in the brain. Bokil et al~\cite{bokil2001ephaptic} use a
simplified model based on cable theory, and find that an action
potential in a single axon can evoke action potentials in neighboring
axons. A more detailed model for coupling intra- and extracellular
currents is the
\emph{Extracellular-Membrane-Intracellular} (EMI)
model~\cite{krassowska1994response, ying2007hybrid,
agudelo2013computationally, agudelo2012numerical,
tveito2017evaluation, tveito2017cell}. The EMI model incorporates
explicit 3D shapes of the neuron, allowing for morphologically
detailed descriptions of the neuropil. However, neither of the
aforementioned frameworks explicitly model the ion concentrations and
can therefore not capture ephaptic effects due to electrodiffusion,
such as diffusive potentials.

The most physically detailed scheme for modelling electrodiffusion is the
\emph{Poisson-Nernst-Planck} (PNP) framework~\cite{lopreore2008computational,
pods2013electrodiffusion, holcman2015new,
cartailler2017electrostatics, cartailler2017analysis}. The PNP
framework is based on explicitly simulating charge relaxation
processes taking place at small spatiotemporal scales ($\sim$nm and
$\sim$ns), and thus requires high resolutions in both time and
space. Consequently, applications have been limited to studying
dynamics at the ion channel and cell membrane level. An alternative
approach is to assume that the bulk tissue is electroneutral, thus
circumventing the need for explicit modelling of charge relaxation
processes. Models based on the electroneutrality assumption are
therefore numerically stable for coarser spatial and temporal
resolutions, allowing for longer simulations on larger
domains. 
 
On this background, a series of electroneutral models for ionic
electrodiffusion have been developed, both for homogenized
domains~\cite{mori2008ephaptic, niederer2013regulation, Halnes2013,
halnes2016, Halnes2017, pods2017comparison, solbra2018kirchoff}, and
for domains including an explicit geometrical representation of the
cells and of the extracellular space~\cite{mori2009numerical}. In
particular, Mori~\cite{mori2009numerical} presents a finite volume
method for solving a system of equations describing cellular
electrical activity accounting for both geometrical effects and ion
concentration dynamics. 

In this paper, we present a variation of the
Mori~\cite{mori2009numerical} model and introduce a mortar-based
finite element formulation of this model. Key advantages of the finite
element formulation are (i) the independence of dimension: the same
scheme is applicable for one-, two- or three-dimensional domains (with
zero-, one- or two-dimensional cell membranes/interfaces); (ii) the
handling of complicated interface geometries; and (iii) the
straightforward use of more accurate i.e.~higher order polynomial
schemes. The framework can be viewed as a combination of the EMI
framework and the electroneutral \emph{Kirchhoff-Nernst-Planck} (KNP)
framework~\cite{solbra2018kirchoff}, and will henceforth be referred
to as the KNP-EMI framework. Previous numerical schemes for the KNP
framework are restricted to simplified 1D geometries~\cite{Halnes2013,
Halnes2015}, or components within a hybrid modelling scheme to compute
extracellular dynamics~\cite{halnes2016, Halnes2017,
solbra2018kirchoff}.

The KNP-EMI framework can be viewed as an extension of the EMI
framework by the explicit modelling of ion concentrations and the
effects of ionic electrodiffusion. We here evaluate the effect of
these extensions by comparing the KNP-EMI and EMI solutions in
idealized axon domains, and find that the solutions are qualitatively
similar but differ locally. However, the KNP-EMI simulations give
further insights into the importance of extracellular bulk conductivities
for ephaptic couplings in neural tissue: KNP-EMI simulations of
idealized, unmyelinated axon bundles reveal increased extracellular
bulk conductivities and, as a result, a reduced tendency toward induction
of action potentials in neighboring axons.
 
\section{Methods}

We present the governing equations for ionic electrodiffusion in
neural tissue with a geometrically explicit representation of the
cellular membranes in Section~\ref{sec:knp-emi} below. To take full
advantage of this framework, a numerical solution scheme capable of
efficiently handling three-dimensional, complicated geometries is
required. We here propose a novel numerical solution scheme using a
mortar finite element method (\cite{bernardi1993domain,
agudelo2013computationally}) and a two-step splitting scheme,
described in Section~\ref{sec:numericalmethods}. This solution
algorithm flexibly allows for arbitrary geometries and efficient
solution of the separate subproblems. Our implementation of this
algorithm is openly available~\cite{CodeZenodoDoi}.

\subsection{A mathematical framework for electrodiffusion with explicit membrane representation}
\label{sec:knp-emi}

\subsubsection{Representation of the computational domain}
\begin{figure}
  \centering \includegraphics[width=1.0\textwidth]{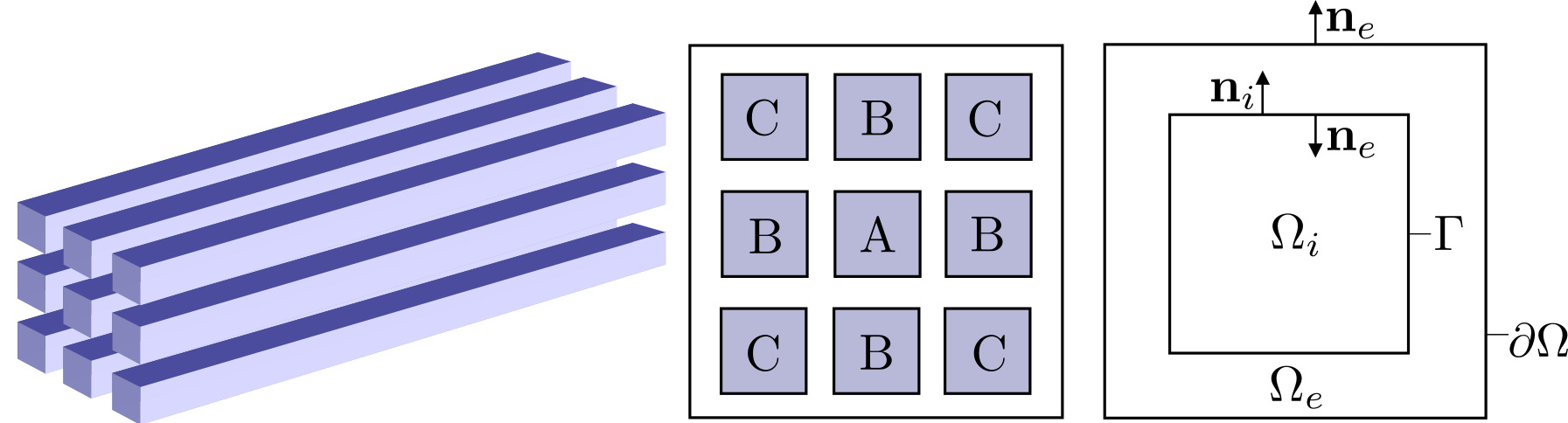}
  \caption{\label{fig:methods} Overview of the computational domains.
    Left: Idealized axon bundle consisting of 9 cuboid-shaped
    axons. Middle: Cross-section of the axon bundle, where the axons
    are labeled with repeated labels for symmetric positions.  Right:
    Idealized 2D computational domain with one intracellular region
    $\Omega_i$ and extracellular region $\Omega_e$.}
\end{figure}

We consider $N$ domains $\Omega_{i^n} \subset \R^d$ ($d = 1, 2, 3$) for $n = 1,
\dots, N$ representing disjoint intracellular regions (physiological cells,
e.g. neurons) and an extracellular region $\Omega_e$, and let the complete
domain $\Omega = \Omega_{i^1} \cup \dots \cup \Omega_{i^N} \cup \Omega_e$ with
boundary $\partial \Omega$. See Figure~\ref{fig:methods} (Right) for an
illustration of a sample domain configuration. We denote the cell membrane
associated with cell $i^n$, i.e.~the boundary of the physiological cell
$\Omega_{i^n}$, by $\Gamma_n$. We assume that $\Gamma_n \cap \Gamma_m =
\emptyset$ for all $n \not = m$ and that $\Gamma_n \cap \partial \Omega =
\emptyset$. (It follows that $\partial \Omega_{i^n} \cap \partial \Omega_e
= \emptyset$ for all $n = 1, \dots, N$.) For simplicity and clarity, we present
the mathematical model for one intracellular region $\Omega_{i^1} = \Omega_{i}$
with membrane $\Gamma$ below. The extension to multiple intracellular regions
is immediate (but notationally cumbersome).

\subsubsection{Intracellular and extracellular governing equations}

We will here derive a system of coupled, time-dependent, nonlinear partial
differential equations to describe ionic electrodiffusion in this domain. We
consider a set of ion species $K$. Typically $K$ will include sodium $\Naion$,
potassium $\Kion$, and chloride $\Clion$. For each ion species $k \in K$ and
each region $r \in \{i, e\}$, we model the \emph{ion concentrations} $[k]_r:
\Omega_r \times (0, T] \rightarrow \R$ (mol/m$^3$) and the \emph{electrical
potentials} $\phi_r: \Omega_r \times (0, T] \rightarrow \R$ (V). Conservation
of ions for the bulk of each region $\Omega_r$ stipulates that
\begin{equation}
  \label{eq:concentration} 
  \frac{\partial [k]_{r}}{\partial t} + \Div \Jvec_{r}^k = 0
  \qquad \text{in } \Omega_{r}, \quad \text{for } r \in \{i, e\}, 
\end{equation}
for $t \in (0, T]$. Here, $\Jvec_r^k: \Omega_r \times (0, T]
    \rightarrow \R^d$ is the regional ion flux density (mol/(m$^2$s))
    of ion $k$. To proceed, we invoke the KNP assumption of bulk
    electroneutrality. In this case, the ion flux densities
    $\Jvec_r^k$ satisfy:
\begin{equation}
  \label{eq:phi}
  - F \sum_{k \in K} z^k \Div \Jvec_{r}^k = 0
  \qquad \text{in } \Omega_{r}, \quad \text{for } r \in \{i, e\}, 
\end{equation}
where $z^k$ is the valence of ion species $k$ and $F$ is Faraday's constant.
The assumption~\eqref{eq:phi} states that the total net flow of ions (weighted
by the respective valences) out of any infinitesimal representative bulk volume
is zero. Furthermore, we assume that the each regional ion flux density can be
expressed by a Nernst-Planck equation as follows:
\begin{equation}
  \label{eq:fluxJ}
  \Jvec_{r}^k = - D_{r}^k \nabla[k]_{r}
  - \frac{D_{r}^k z^k}{\psi}[k]_{r} \nabla \phi_{r},
  \qquad \text{in } \Omega_r, \quad r \in \{i, e\}.
\end{equation}
Here, $D_r^k$ denotes the effective diffusion coefficient (m$^2$/s) of ion
species $k$ in the region $r$. The constant $\psi = R T F^{-1}$ combines
Faraday's constant $F$, the absolute temperature $T$, and the gas constant $R$.
The ion flux density, i.e.~the flow rate of ions per unit area, is thus
modelled as the sum of two terms: (i) the diffusive movement of ions due to
ionic gradients $- D_r^k \Grad [k]_r$ and (ii) the ion concentrations that are
transported via electrical potential gradients, i.e.~the ion migration $- D_r^k
z^k \psi^{-1} [k]_r \nabla \phi_r$ where $D_r^k \psi^{-1}$ is the
electrochemical mobility. This model ignores convective effects, and thus
assumes that the underlying material (typically fluid) is at rest.  As the
potential $\phi_e$ is only determined up to a constant in
equations~\eqref{eq:concentration}--\eqref{eq:fluxJ}, an additional constraint
is required,~e.g:
\begin{equation}
    \label{eq:constraint}
    \int_{\Omega_e} \phi_e \dx = 0.
\end{equation}

By inserting~\eqref{eq:fluxJ} into~\eqref{eq:phi} we recognize (from volume
conductor theory) the following expression for the bulk conductivity $\sigma_r$:
\begin{equation}
    \label{eq:sigmar} \sigma_r =
    \frac{F}{\psi}\sum_{k \in K} D_r^k[k]_r(z^k)^2.
\end{equation}
Notably, the bulk conductivity $\sigma_r$ depends on the ion concentrations $[k]_r$
and the diffusion coefficients $D_r^k$.  Electrodiffusive models without
explicit modelling of the ion concentrations typically set the bulk conductivity as
a independent parameter, e.g.~\cite{krassowska1994response, tveito2017cell,
bokil2001ephaptic}.

Inserting~\eqref{eq:fluxJ} into~\eqref{eq:concentration} and
into~\eqref{eq:phi}, we thus obtain a system of $N|K| + N$ equations ($N|K|$
parabolic, $N$ elliptic) for the $N|K| + N$ unknown scalar fields. The system
remains to be closed by appropriate initial conditions, boundary conditions,
and importantly interface conditions.

\subsubsection{Interface conditions}

We next turn to modelling the cell membrane currents and membrane potential
across the interface $\Gamma$. We denote the membrane potential $\phi_M$ as the
jump in the electrical potential over the membrane:
\begin{equation}
  \label{eq:mem:potentialrel}
  \phi_M = \phi_{i} - \phi_e \qquad \text{ on } \Gamma. 
\end{equation}

We introduce the \emph{total ionic current density} $I_{M}: \Gamma \times (0,
T) \rightarrow \R$ (C/(m$^2$s)) across the interface $\Gamma$. By definition
and by conservation of total charge, we have that
\begin{equation}
  \label{eq:membrane:conservation}
  I_M \equiv - F \sum_{k \in K} z^k \Jvec_e^k \cdot \nvec_e
  =  F \sum_{k \in K} z^k \Jvec_i^k \cdot \nvec_i .
\end{equation}
where $\nvec_r$ denotes the boundary normal pointing out of $\Omega_r$
for $r \in \{i, e\}$. Next, we assume that $I_M$ consists of two
components: (i) a total channel current $\Ich$ and (ii) a capacitive
current $\Icap$:
\begin{equation}
  \label{eq:mem:totCur}
  I_M = \Ich + \Icap .
\end{equation}
We further assume that the total channel current $\Ich$ is the sum of the ion
specific channel currents $\Ich^k$:
\begin{equation}
  \label{eq:channel-currents}
  \Ich = \sum_{k \in K} \Ich^k, \quad \Ich^k = \Ich^k(\phi_M, [k]_{\cdot}, ...).
\end{equation}
The channel currents $\Ich^k$ are subject to modelling. Typical models for
$\Ich^k$ notably includes an synaptic input current $\Isyn$, leaky passive
neuron, Hodgkin-Huxley etc, and will be detailed further below in
Section~\ref{sec:mem}.  On the other hand, the capacitive current $\Icap$ is
defined over to be the capacitance $C_M$ times the rate of change of the
voltage~\cite{sterratt2011principles}, hence:
\begin{equation}
  \label{eq:mem:defCapCur}
  \Icap = C_M\frac{\partial \phi_M}{\partial t} .
\end{equation}
Inserting~\eqref{eq:mem:defCapCur} into~\eqref{eq:mem:totCur} and
rearranging gives the following relation for the membrane potential
$\phi_M$:
\begin{equation}
  \label{eq:mem:potential}
  \frac{\partial \phi_M}{ \partial t} = \frac{1}{C_M}(I_M - I_\text{ch}) .
\end{equation}

It remains to specify a set of interface conditions for the specific ion fluxes
$\Jvec_r^k \cdot \nvec_r$ for $r \in \{i, e\}$. Here, we propose a heuristic
approach via ion specific capacitive current modelling. An alternative approach
is presented in~\cite{mori2009numerical}. As for the total current, we assume
that the capacitive current can be represented as a sum of ion-associated
currents:
\begin{equation}
  \Icap = \sum_{k \in K} \Icap^k .
\end{equation}
Without loss of generality, we let the ion specific capacitive current
$\Icapr^k$ in region $\Omega_r$ at the interface $\Gamma$ be some fraction
$\alpha_r^k$ of the total capacitive current $\Icap$:
\begin{equation}
  \label{eq:rel:capCurIon}
  \Icapr^k = \alpha_r^k \Icap .
\end{equation}
Specifically, we assume that:
\begin{equation}
  \label{eq:mem:alpha}
  \alpha_r^k = \frac{ D_r^k (z^k)^2 [k]_{r} }{\sum_{l\in K}  D_r^l (z^l)^2 [l]_{r}},
\end{equation}
and note that $\sum_{k\in K} \alpha_r^k = 1$ for $r \in \{i, e\}$. By
definition of the ion currents and the expression for the capacitive current
given by~\eqref{eq:mem:totCur}, we let the intracellular and extracellular ion
fluxes across the membrane be given by:
\begin{align}
  \label{eq:mem:conc}
  \Jvec_i^k \cdot \nvec_i = 
  \frac{\Ich^k + \alpha_i^k (I_M - \Ich) }{F z^k}, \qquad
  - \Jvec_e^k \cdot \nvec_e = 
  \frac{\Ich^k + \alpha_e^k (I_M - \Ich) }{F z^k}, 
\end{align} 
for $k \in K$.

\subsubsection{Modelling specific ion channels}
\label{sec:mem}

The framework presented thus far allows for general representations of the ion
channel current dynamics. In particular, the framework admits different choices
of ion specific channel current models $\Ich^k$. An advantage of the
geometrically explicit framework is that it allows for different channel
currents models for individual cells and e.g.~geometrically heterogeneous
material properties. We here summarize two examples of ion specific channel
currents: a passive membrane model~\cite{sterratt2011principles} and the
Hodgkin-Huxley model~\cite{hodgkin1952quantitative}.

\paragraph{Passive membrane dynamics}
We model the passive membrane channel current for ion species $k$
as~\cite{sterratt2011principles}:
\begin{equation}
  \label{eq:passive}
  \Ich^k(\phi_M) = g_L^k(\phi_M - E^k),
\end{equation}
where $g_L^k$ is a constant leak conductivity, and $E^k$ is the ion
specific reversal potential, given by
\begin{equation*}
  E^k = \frac{RT}{z^k F}\ln \frac{[k]_e}{[k]_i},
\end{equation*}
with valence $z^k$, Faraday's constant $F$, absolute temperature $T$,
and gas constant $R$.

\paragraph{Hodgkin-Huxley membrane dynamics}
In order to model active membrane dynamics, we use the standard Hodgkin-Huxley
membrane model~\cite{hodgkin1952quantitative}. The ion species under
consideration are sodium \Naion, potassium \Kion, and chloride \Clion, and the
model additionally introduces three gating variables $m, h, n$ associated with
sodium channel activation, potassium channel activation and potassium channel
inactivation, respectively. The membrane potential $\phi_M$ is then modelled by
the following specialization of~\eqref{eq:mem:potential}:
\begin{equation}
  \label{eq:HH:potential}
  \frac{\partial \phi_M}{ \partial t} = \frac{1}{C_M}(I_M - \Ich^{\Na} -
    \Ich^{\K} - \Ich^{\Cl}),
\end{equation}
with ion specific membrane channel currents:
\begin{align}
  \label{eq:HH:currents}
  \Ich^{\Na}(\phi_M) &= \bar{g}^{\Na} m^3 h(\phi_M - E^{\Na}),  \\
  \Ich^{\K}(\phi_M) &= \bar{g}^{\K} n^4 (\phi_M - E^{\K}),  \\
  \Ich^{\Cl}(\phi_M) &= \bar{g}^{\Cl} (\phi_M - E^{\Cl}).
\end{align}
Here, $\bar{g}^k$ is the maximal conductivity for ion species $k$. The
gating variables are governed by the following ODE:
\begin{equation}
    \label{eq:HH:ODEs}
  \frac{\partial p}{\partial t} = \alpha_p(\phi_M)(1-p) - \beta_p(\phi_M)p,
\end{equation}
for $p \in \{m,h,n\}$. The rate constants $\alpha_p$ and $\beta_p$ take the form
\begin{align}
    \label{eq:HH:ratecoefficients:alpha}
  \alpha_p(\phi_M) &= p_{\infty}(\phi_M)/\tau_p, \\
    \label{eq:HH:ratecoefficients:beta}
  \beta_p(\phi_M) &= (1 - p_{\infty}(\phi_M))/\tau_p,
\end{align}
where $p_{\infty}$ is the steady state value for activation and
$\tau_p$ is the time constant.

\subsubsection{Initial and boundary conditions}

We assume that initial conditions are given for all ion concentrations, both
intracellularly and extracellularly:
\begin{equation}
  \label{eq:initial:k}
  [k]_r(x, 0) = [k]_r^0(x) \qquad x \in \Omega_r, \quad r \in \{i, e\}.
\end{equation}
Furthermore, we assume that these conditions are compatible with the assumption
of bulk electroneutrality, i.e. that the initial state of the system satisfies:
\begin{equation}
  \sum_{k\in K} z^k [k]^0_e = 0.
\end{equation}
In addition, we assume that an initial condition is given for the
membrane potential:
\begin{equation}
  \label{eq:initial:phi}
  \phi_M(x, 0) = \phi_M^0(x), \qquad x \in \Gamma.
\end{equation}

Finally, a set of boundary conditions will close the system. We describe
specific boundary conditions in the numerical experiments in
Section~\ref{sec:results}.

\subsubsection{Summary of governing equations}

In summary, the mathematical framework for electrodiffusion with
explicit geometrical representation of the cell membranes is comprised
of the bulk equations~\eqref{eq:concentration}, \eqref{eq:phi}
with~\eqref{eq:fluxJ}, the interface
conditions~\eqref{eq:membrane:conservation},~\eqref{eq:mem:potential}
with~\eqref{eq:mem:potentialrel} and \eqref{eq:channel-currents},
and~\eqref{eq:mem:conc} with \eqref{eq:mem:alpha}, the initial
conditions~\eqref{eq:initial:k} and \eqref{eq:initial:phi}, and
additional boundary conditions. We will refer to this set of equations
as the KNP-EMI framework.

\subsection{Numerical methods}
\label{sec:numericalmethods}

To solve the KNP-EMI framework numerically, we consider a finite difference
time integration scheme, a splitting scheme, and a mortar finite element method
in space. We derive the new finite element scheme and describe the splitting
algorithm in the sections below.

\subsubsection{Weak formulation of the governing equations}
\label{sec:weak}
Multiplying~\eqref{eq:concentration} with test functions $v_r^k$ (for
$r \in \{i, e\}$), integration over the intracellular and
extracellular domains $\Omega_i$ and $\Omega_e$ separately,
integration by parts, and inserting~\eqref{eq:mem:conc} for the ion
fluxes across the membrane, yields
\begin{align}
  \label{eq:num:1}
  \int_{\Omega_i} \frac{\partial [k]_i}{\partial t} v_i^k
  - \Jvec_i^{k} \cdot \nabla v_i^k \dx
  + \frac{1}{F z^k} \int_{\Gamma} \left (\Ich^k + \alpha_i^k \left(I_{M} - \Ich \right) \right )v^k_i \ds 
  &= 0, \\
  \label{eq:num:2}
  \int_{\Omega_e} \frac{\partial [k]_e}{\partial t} v_e^k 
  - \Jvec_e^{k} \cdot \nabla v_e^k \dx
  - \frac{1}{F z^k} \int_{\Gamma} \left (\Ich^k + \alpha_e^k \left(I_{M} - \Ich \right) \right ) v^k_e \ds 
  &= 
  - \int_{\partial \Omega} \Jvec_e^k \cdot \nvec_e \, v_e^k \ds.
\end{align}
Similarly, multiplying~\eqref{eq:phi} by test functions $w_r$ for $r
\in \{i, e\}$, integration by parts and
inserting~\eqref{eq:membrane:conservation} for the total membrane
current, yields
\begin{align}
  \label{eq:num:3}
  F \sum_{k \in K} z^k \int_{\Omega_i} \Jvec_{i}^{k} \cdot \Grad w_i \dx 
  - \int_{\Gamma} I_M \, w_i \ds
  &= 0, \\
  \label{eq:num:4}
  F \sum_{k \in K} z^k \int_{\Omega_e} \Jvec_{e}^{k} \cdot \Grad w_e \dx
  + \int_{\Gamma} I_M \, w_e \ds
  &= F \sum_{k \in K} z^k \int_{\partial \Omega} \Jvec_e^{k} \cdot \nvec_e \, w_e
    \ds.
\end{align}
The constraint~\eqref{eq:constraint} is enforced by introducing an additional
unknown (a Lagrange multiplier) $c_e \in \R$ along with a test function $d_e
\in R$, in the following manner:
\begin{equation}
    \int_{\Omega_e} c_e w_e \dx = -\int_{\Omega_e} \phi_e d_e \dx.
\end{equation}
Finally, multiplying~\eqref{eq:mem:potential} by a test function $q$,
and integrating over $\Gamma$ yields
\begin{equation}
  \label{eq:num:5}
  C_M \int_{\Gamma} \frac{\partial (\phi_i - \phi_e)}{\partial t} \, q \ds -
    \int_{\Gamma} (I_M - \Ich) \, q \ds = 0.
\end{equation}

We remark that this is a weak formulation of a set of time-dependent, nonlinear
equations. In particular, recall that $\Ich$ and $\Ich^k$ depend on $\phi_M$
and $[k]_r$ cf.~\eqref{eq:channel-currents} while $\alpha_r^k$ depends on
$[k]_r$ cf.~\eqref{eq:mem:alpha}.

To solve this system numerically, we consider the following
approximations.
\begin{itemize}
\item We discretize the time derivatives in~\eqref{eq:num:1}--\eqref{eq:num:2}
and~\eqref{eq:num:5} using a finite difference method.
\item We approximate $\Jvec_r^k$ at time $t^n$ by the linearized ion
    flux density
(cf.~\eqref{eq:fluxJ}):
\begin{equation*}
    \Jvec_r^{k} \approx - D_r^k
    \nabla[k]^{n}_r - \frac{D_r^k z^k}{\psi}[k]^{n-1}_i \nabla \phi^{n}_i.
\end{equation*} 
\item We evaluate $\alpha_r^k$ at time $t^n$ by the previous
value (cf.~\eqref{eq:mem:alpha}):
\begin{equation} 
  \label{eq:approx:alpha}
  \alpha_r^k \approx \frac{ D_r^k (z^k)^2 [k]_{r}^{n-1} }{\sum_{l \in
      K} D_r^l (z^l)^2 [l]^{n-1}_{r}}.
\end{equation}
\end{itemize}
Moreover, we evaluate $\Ich$ and the discretization
of~\eqref{eq:num:5} depending on the choice of ion channel model
(cf.~Section~\ref{sec:mem}) as follows.
\begin{itemize}
  \item
    For the passive model, we insert the linear relation given
    by~\eqref{eq:passive} directly
    in~\eqref{eq:num:1}--\eqref{eq:num:2}
    and~\eqref{eq:num:5}. Moreover, the implicit discretization
    of~\eqref{eq:num:5} reads as:
    \begin{equation}
      \frac{\partial (\phi_i - \phi_e)}{\partial t} \approx 
        \Delta t^{-1}(\phi_M^n - \phi_M^{n-1}),
    \end{equation}
    at time $t^n$ with $\phi_M^{n} = \phi_i^n - \phi_e^n$ and $\Delta t = t^n -
        t^{n-1}$.
  \item
    For the Hodgkin-Huxley model, we use the following two-step
    splitting procedure. Consider $n \in [1, \dots, N]$ with $t^n -
    t^{n-1} = \Delta t$, and assume that $[k]_r^{n-1}$
          and $\phi_{M}^{n-1}$ at time step $t^{n-1}$ are known.
  \begin{itemize}
  \item In the first (ODE) step, we update the membrane potential
    $\phi_M^{n}$ at time step $t^n$ by solving the ODE
    system~\eqref{eq:HH:potential}--\eqref{eq:HH:ratecoefficients:beta},
    with $I_M$ set to zero, using $25$ explicit (forward) Euler steps
    of size $\Delta t^* = \Delta t/25$.

  \item In the second (PDE) step, we solve for $[k]_r^n$, $\phi_r^n$
    and $I_M^n$ (for $r \in \{i, e\}$) in the linear system arising
    from spatial discretization of~\eqref{eq:num:1}--\eqref{eq:num:5},
    with $\Ich$ set to zero in~\eqref{eq:num:5}, and $\Ich$
    approximated by
    \begin{equation}
      \label{eq:approx:Ich}
      \Ich \approx \Ich(\phi_M^{n}, [k]_{\dots}^{n-1}),
    \end{equation}
    in~\eqref{eq:num:1}--\eqref{eq:num:2}, where $\phi_M^{n}$ is the
    membrane potential solution at $t^n$ from the ODE step (see
    Section~\ref{sec:fem} for details). The implicit discretization
    of~\eqref{eq:num:5} reads as:
    \begin{equation}
      \label{eq:num:approx:1}
      \frac{\partial (\phi_i - \phi_e)}{\partial t} \approx  \Delta t^{-1}(\phi_M^n -
      \phi_M^{n-1}),
    \end{equation}
    where $\phi_M^{n} = \phi_i^n - \phi_e^n$ is the membrane potential solution at $t^n$ from
    the ODE step.
  \end{itemize}
\end{itemize}
The steps are repeated until global end time $t^N$ is reached.

\subsubsection{Spatial discretization}
\label{sec:fem}

\begin{figure}
  \centering \includegraphics[width=1.0\textwidth]{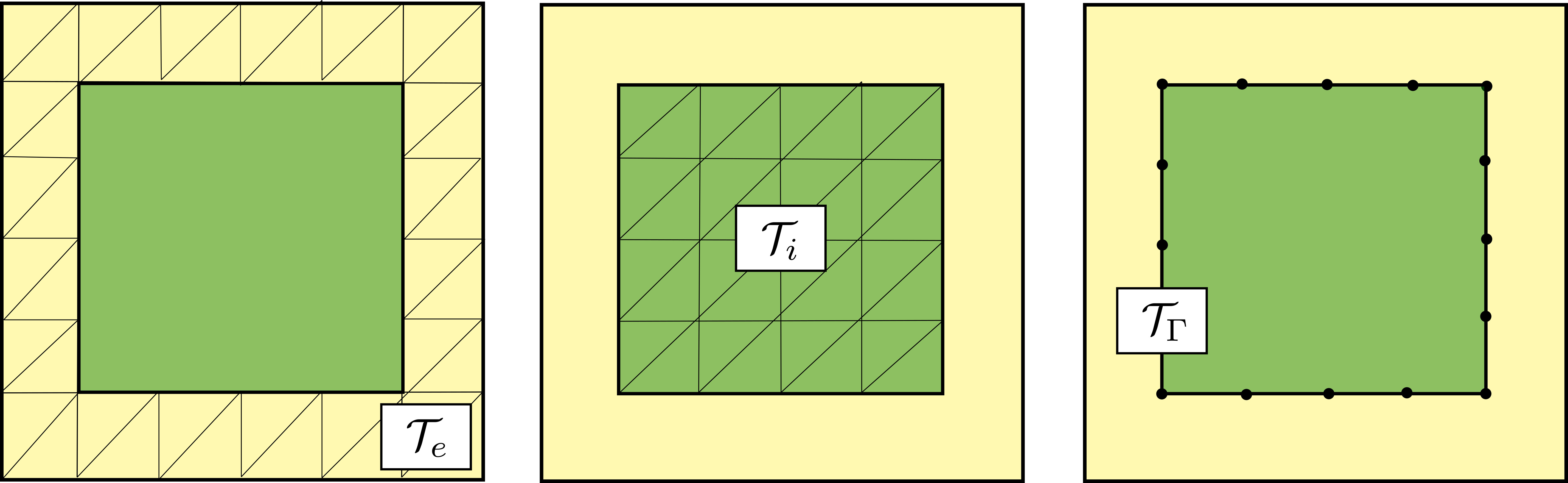}
  \caption{Schematic representation of meshes for the discretization
    of the PDE part of the governing electrodiffusive equations using
    a mortar finite element method. Mesh $\mathcal{T}_e$ of the
    extracellular subdomain $\Omega_e$ (left), mesh $\mathcal{T}_i$ of
    the intracellular subdomain $\Omega_i$ (middle) and mesh
    $\mathcal{T}_{\Gamma}$ of the interface $\Gamma$ (right). Note
    that the shared facets of the extracellular and intracellular
    meshes form the (codimension 1) mesh of the interface.}
  \label{fig:mortar}
\end{figure}

To numerically solve the PDE part of the governing equations defined
on the domain $\Omega = \Omega_i \cup \Omega_e$, we use a mortar
finite element method. We discretize each subdomain $\Omega_r$ by a
conforming mesh $\mathcal{T}_r$ for $r \in \{i, e\}$. We assume that
the meshes $\mathcal{T}_i$ and $\mathcal{T}_e$ match at the common
interface $\Gamma$, and define a (lower-dimensional) mesh
$\mathcal{T}_{\Gamma}$ of this interface
(cf.~Figure~\ref{fig:mortar}).

Next, we introduce separate finite element spaces for approximating the unknown
fields in the weak formulation~\eqref{eq:num:1}--\eqref{eq:num:5}, $[k]_r :
\Omega_r \rightarrow \R, \phi_r : \Omega_r \rightarrow \R$ for $r \in \{i,
e\}$, $c_e : \Omega_e \rightarrow \R$, and $I_m : \Gamma \rightarrow \R$. We
approximate the ion concentrations $[k]_r$ and potentials $\phi_r$ using
continuous piecewise linear polynomials (linear Lagrange finite elements) over
the meshes $\mathcal{T}_r$. These fields thus have degrees of freedom defined
on the vertices of the extracellular and intracellular meshes. The Lagrange
multiplier $c_e$ is approximated using a single real number. Furthermore, the
transmembrane current $I_M$ is approximated using continuous piecewise linear
polynomials over the facet mesh $\mathcal{T}_{\Gamma}$. We denote the finite
element spaces for approximating $[k]_r$ by $V_r^k$, the spaces for
approximating $\phi_r$ by $W_r$ and the spaces for approximating $I_M$ by $Q$.
Let $\inner{u}{v}_{\Omega} = \int_{\Omega} u v \dx$. For notational simplicity,
we denote the approximation of $[k]_r$ by $[k]_r$, the approximation of
$\phi_r$ by $\phi_r$, and the approximation of $I_M$ by $I_M$ below. We here
use linear polynomials for concreteness, but the formulation also applies
directly for higher order polynomials.

We then solve the PDE step in the two-step splitting scheme described
in Section~\ref{sec:weak} as follows: given $[k]_{r}^{n} \in
V_{r}^k$ and $\phi_{M}^{n} \in Q$ at time step $t^{n}$, and the
previously computed $\Ich^k$ and $\alpha^k$
(cf.~\eqref{eq:approx:Ich} and~\eqref{eq:approx:alpha}), find the ion
concentration $[k]_{r} \in V_r^k$, the potential $\phi_{r} \in
W_{r}$, and the total transmembrane current density ${I}_{M} \in Q$
at time step $t^{n+1}$ such that:
\begin{align*}
    \label{eq:weakform:discrete}
    \frac{1}{\Delta t}\inner{[k]_{i}}{ v_i^k}_{\Omega_i}
    -  \inner{ \Jvec_{i}^k}{\nabla v_i^k }_{\Omega_i}
    - \inner {\alpha_i^k I_{M}}{v_i^k}_{\Gamma} &=
    \frac{1}{\Delta t}\inner{[k]_{i}^n}{ v_i^k}_{\Omega_i} 
    + \inner{\frac{\Ich^k - \alpha_i^k \Ich}{Fz^k}}{ v^k_i}_{\Gamma}, \\
    \frac{1}{\Delta t}\inner{[k]_{e}}{v_e^k}_{\Omega_e} 
    -  \inner{ \Jvec_{e}^k}{ \nabla v_e^k}_{\Omega_e}
    + \inner{ \alpha_e^k\ I_{M}}{ v_e^k}_{\Gamma} &= 
    \frac{1}{\Delta t}\inner{ [k]_{e}^n}{v_e^k}_{\Omega_e}
    - \inner {\frac{\Ich^k - \alpha_e^k \Ich}{Fz^k}}{ v^k_e}_{\Gamma}
    - \inner{ \Jvec_{e}^k \cdot \nvec_e}{v_e^k}_{\partial \Omega} , \\
    F \sum_{k \in K} z^k \inner{ \Jvec_i^{k}}{ \nabla w_i}_{\Omega_i}
    + \inner{ I_{M}}{w_i}_{\Gamma}  &= 0,  \\
    F \sum_{k \in K} z^k \inner{ \Jvec_e^{k}}{\nabla w_e}_{\Omega_e}
    - \inner{I_{M}}{ w_e}_{\Gamma}
    &= F \sum_{k \in K} z^k \inner{ \Jvec_{e}^k\cdot\nvec_e}{w_e}_{\partial \Omega}, \\
    \inner{c_e}{w_e}_{\Omega_e} + \inner{\phi_e}{d_e}_{\Omega_e} &= 0, \\
    \frac{1}{\Delta t} \inner{ \phi_{i} - \phi_{e}}{q}_{\Gamma}
    - \frac{1}{C_M} \inner{ I_{M}}{q}_{\Gamma}
    &=  \frac{1}{\Delta t} \inner{ \phi^n_{M}}{ q}_{\Gamma},
\end{align*}
for all $v_i^k \in V_i^k$, $v_e^k \in V_e^k$, $w_i \in W_i$, $w_e \in
W_e$, $d_e \in D_e$ and $q \in Q$. The ion flux terms on the right-hand side are
replaced by appropriate boundary conditions in the subsequent
sections.

To evaluate the accuracy of the numerical solutions defined over
$\Omega_r$ for $r \in \{i, e\}$, we use the standard $L^2$ and $H^1$
norms denoted by $\| \cdot \|_0$ and $\| \cdot \|_1$, respectively: for $u :
\Omega_r \rightarrow \R$,
\begin{equation*}
  \| u \|_{0}^2 = \int_{\Omega_r} u^2 \dx, \quad
  \| u \|_{1}^2 = \int_{\Omega_r} u^2 + \Grad u \cdot \Grad u \dx.
\end{equation*}
In addition, for $I : \Gamma \rightarrow \R$, we define the broken
$L^2$-norm by summing over the $L^2$-norms over the mesh cells of the
interface mesh $\mathcal{T}_{\Gamma}$:
\begin{equation*}
  \| I \|_{0, \Gamma}^2 = \sum_{f \in \mathcal{T}_{\Gamma}} \| I |_{f} \|_0^2.
\end{equation*}

\subsubsection{Implementation}
The numerical scheme was implemented using a mixed dimensional
framework from the FEniCS finite element
library~\cite{alnaes2015fenics}. The linear systems arising in the
numerical experiments were solved using a direct (MUMPS) solver.
The code is publicly available~\cite{CodeZenodoDoi}.

\subsubsection{Comparison with EMI framework}
In the numerical experiments comparing the KNP-EMI and the EMI
models, the EMI model is discretized using the mortar finite
element formulation as presented in Tveito et
al~\cite{tveito2017evaluation}.

\subsection{Computational models and parameters}
\label{sec:parameters}
We consider two model set-ups
for testing the presented methodology (Model A and B), a model (Model
C) for comparing simulation results between the KNP-EMI and EMI
frameworks, and a model for studying ephaptic coupling (Model D). The
model set-ups are described in detail here. The model parameters are
given in Table \ref{tab:physconst}, unless otherwise stated in the
text. We assume that all axons in each simulation have the same
membrane channel current $\Ich$. We denote the spatial coordinates in
this and subsequent sections by $(x, y, z)$.
\begin{table}
    \begin{center}
    \begin{tabular}{lllll}
    \hline
        Parameter & Symbol & Value & Unit & Reference \\
    \hline
        gas constant       & $R$ &  8.314             & J/(K mol)   & -- \\
        temperature        & $T$ &  300               & K           & --\\
        Faraday's constant & $F$ &  $9.648\cdot 10^4$ & C/mol       & --\\ 
        membrane capacitance & $C_M$ & $1 \cdot 10^{-5}$ & nF/\textmu m & \\ 
        Na\textsuperscript{+} diffusion coefficient & $D^\text{Na}_r$ & 1.33 &\textmu m\textsuperscript{2}/ms & \cite{hille2001ion}\\
        K\textsuperscript{+} diffusion coefficient  & $D^\text{K}_r$  & 1.96 &\textmu m\textsuperscript{2}/ms & \cite{hille2001ion}\\
        Cl\textsuperscript{\textminus} diffusion coefficient & $D^\text{Cl}_r$ &  2.03 & \textmu m\textsuperscript{2}/ms & \cite{hille2001ion}\\
        Na\textsuperscript{+} leak conductivity & $g^\text{Na}_L$ & 0.2 & mS/cm\textsuperscript{2}  & \\
        K\textsuperscript{+} leak conductivity & $g^\text{K}_L$  & 0.8 & mS/cm\textsuperscript{2} & \\ 
        Cl\textsuperscript{-} leak conductivity & $g^\text{Cl}_L$ & 0   & mS/cm\textsuperscript{2} & \\
        K\textsuperscript{+}  HH max conductivity & $\bar{g}^\text{K}$ &36  & mS/cm\textsuperscript{2} & \cite{hodgkin1952quantitative} \\
        Na\textsuperscript{+} HH max conductivity & $\bar{g}^\Na$   & 120 & mS/cm\textsuperscript{2} & \cite{hodgkin1952quantitative} \\
        synaptic time constant & $\alpha$ &  1 & ms & \\
        initial intracellular Na\textsuperscript{+} concentration & $[\text{Na}]_i^0$ & 12 & mM & \cite{pods2013electrodiffusion} \\ 
        initial extracellular Na\textsuperscript{+} concentration & $[\text{Na}]_e^0$ & 100& mM & \cite{pods2013electrodiffusion} \\ 
        initial intracellular K\textsuperscript{+} concentration  & $[\text{K}]_i^0$ & 125 & mM & \cite{pods2013electrodiffusion}\\
        initial extracellular K\textsuperscript{+} concentration &$[\text{K}]_e^0$ & 4 & mM & \cite{pods2013electrodiffusion}\\
        initial intracellular Cl\textsuperscript{\textminus} concentration & $[\text{Cl}]_i^0$ & 137 & mM & \cite{pods2013electrodiffusion}\\ 
        initial extracellular Cl\textsuperscript{\textminus} concentration & $[\text{Cl}]_e^0$ & 104 & mM & \cite{pods2013electrodiffusion}\\
        initial membrane potential & $\phi_M^0$ & -67.74 & mV &  \\
        initial HH gating value (Na\textsuperscript{+} activation) & m$^0$ & 0.0379 & -- & \cite{hodgkin1952quantitative}\\
        initial HH gating value (Na\textsuperscript{+} inactivation & h$^0$ & 0.688 & -- & \cite{hodgkin1952quantitative}\\
        initial HH gating value (K\textsuperscript{+} activation) & n$^0$ & 0.276 & -- & \cite{hodgkin1952quantitative}\\
        global time step & $\Delta t$ & 0.1 & ms & -- \\
        local time step & $\Delta t^*$ & $\Delta t/25$ & ms & -- \\
   \hline
    \end{tabular}
    \caption{\label{tab:physconst} The physical parameters and initial values
        used in the simulations. The values are collected from Hille et
        al~\cite{hille2001ion}, Hodgkin et al~\cite{hodgkin1952quantitative},
        and Pods et al~\cite{pods2013electrodiffusion}.} \end{center}
\end{table}

\subsubsection{Model A: One axon with a passive membrane model}
\label{sec:modelA}
For Model A, we consider a two-dimensional domain $\Omega = \Omega_i \cup
\Omega_e = [0, 60] \times [0, 60]$ $\mu$m, with one intracellular domain
(cell) $\Omega_i = [6, 56] \times [28, 34]$ $\mu$m. We mesh this domain by
dividing the domain into $n \times m$ rectangles, with $\Delta x = 60/n$ and
$\Delta y = 30/m$, and dividing each rectangle into two triangles by a
diagonal, for a series of $\Delta x = \Delta y = 2, 1, 0.5, 0.25$ $\mu$m. We
model $\Ich$ using the passive model, as described in Section~\ref{sec:mem},
and prescribe a synaptic input $I_{\rm syn}$ of the form
\begin{equation}
  \label{eq:synapse}
  I_{\rm syn}= g_{\rm syn} H(x) e^{\frac{t-t_0}{\alpha}}(\phi_M - E_\text{Na}),
\end{equation}
where $\alpha$ is the synaptic time constant, $H(x) = \{ 1 \text{ for } x \in Z
\text{ and } 0 \text{ elsewhere} \}$ for an interval $Z$. We let $Z = [5, 10]
\, \mu$m, and set $t_0 = 0$, $g_{\rm syn} = 125$ mS/cm$^2$. At the exterior
boundary $\partial \Omega$, we apply the boundary conditions
\begin{equation}
    \label{eq:bcs:dirichlet}
    [k]_e = [k]_e^0,
    \,  \text{at } \partial \Omega, \quad
    \sum_{k \in K} z^k \Jvec_e^k \cdot \nvec_e = 0,
    \,  \text{at } \partial \Omega,
\end{equation}
describing the extracellular ion concentrations at the exterior boundary,
and that no charge can leave or enter the system.

\subsubsection{Model B: One axon with a passive membrane model and non-physical parameters}
To evaluate the numerical accuracy of the mortar finite element scheme
presented in Section~\ref{sec:numericalmethods}, we construct an analytical
solution using the method of manufactured
solutions~\cite{roache1998verification}. In particular, we let the analytical
solution to \eqref{eq:concentration}--\eqref{eq:mem:conc} be given by:
\begin{equation}
\begin{aligned}
    \label{eq:MMS}
    \text{[Na]}_{i}^e &= 0.7 + 0.3 \sin(2\pi x) \sin(2\pi y)\exp(-t),
    && \qquad \text{in } \Omega_{i}, \\ 
    \text{[Na]}_e^e &= 1.0 + 0.6 \sin(2\pi x) \sin(2\pi y)\exp(-t),
    && \qquad \text{in } \Omega_{e}, \\ 
    \text{[K]}_i^e &= 0.3  + 0.3 \sin(2\pi x) \sin(2\pi y)\exp(-t),
    && \qquad \text{in } \Omega_{i}, \\ 
    \text{[K]}_e^e &= 1.0  + 0.2 \sin(2\pi x) \sin(2\pi y)\exp(-t),
    && \qquad \text{in } \Omega_{e}, \\ 
    \text{[Cl]}_i^e &= 1.0 + 0.6 \sin(2\pi x) \sin(2\pi y)\exp(-t),
    && \qquad \text{in } \Omega_{i}, \\ 
    \text{[Cl]}_e^e &= 2.0 + 0.8 \sin(2\pi x) \sin(2\pi y)\exp(-t),
    && \qquad \text{in } \Omega_{e}, \\ 
    \phi_i^e &= \cos(2\pi x)\cos(2\pi y)(1 +\exp(-t)),
    && \qquad \text{in } \Omega_{i}, \\ 
    \phi_e^e &= \cos(2\pi x)\cos(2\pi y),
    && \qquad \text{in } \Omega_{e},
\end{aligned}
\end{equation}
with the passive model $\Ich = \phi_M$ and with $\Isyn = 0$. We assume that the
parameter values all equal one: $C_m = D_i^k = D_e^k = F = G = R = 1$, and that
$K = \{\Na^+,~\K^+,~\Cl^-\}$.  We consider a two-dimensional domain $\Omega =
\Omega_i \cup \Omega_e = [0, 1] \times [0, 1]$, with one intracellular domain
$\Omega_i = [0.25, 0.75 ] \times [0.25, 0.75]$. The domain is meshed as for
Model A (cf.~Section~\ref{sec:modelA}) for a series of $n = m = 16, 32, 64,
128, 256$. In the numerical experiments for this test case, we initially let
$\Delta t = \frac{1}{64} \times 10^{-5}$, and then quarter the timestep in each
series. The errors are evaluated at $t = \frac{2}{64} \times 10^{-5}$.

\subsubsection{Model C: Multiple axons with a passive membrane model}
\label{sec:model:C}

For Model C, we define three different two-dimensional domains: (C1) a
domain with one intracellular region (cell), (C2) a domain with two
intracellular regions with a distance of 4 $\mu$m in the $y$-direction
between the cells, and (C3) a domain with two intracellular regions
with a distance of 10 $\mu$m in the $y$-direction between the
cells. More precisely, we let
\begin{itemize}
\item[\bf Model C1: ]
  $\Omega = \Omega_i \cup \Omega_e = [0, 120] \times [0,
  120]$ $\mu$m, $\Omega_i = [35, 85] \times [57, 63]$ $\mu$m.
\item[\bf Model C2: ] 
  $\Omega = \Omega_i^1 \cup \Omega_i^2 \cup \Omega_e = [0,120] \times
  [0,120]$ $\mu$m, with two cells $\Omega_i^1 = [35, 85] \times
  [52,58]$ $\mu$m and $\Omega_i^2 = [35,85] \times [62,68]$ $\mu$m.
\item[\bf Model C3: ]
  $\Omega$ as in Model C2 but with $\Omega_i^1 = [35, 85] \times [49,
  55]$ $\mu$m and $\Omega_i^2 = [35,85] \times [65,71]$ $\mu$m. 
\end{itemize}

The ion channel currents $\Ich^k$ are modelled using the passive model
described in Section~\ref{sec:mem}. The synaptic input current
model~\eqref{eq:synapse} is applied with $g_{\text{syn}} = 12.5$
mS/cm\textsuperscript{2}, $t_0=0$, and with $Z = [35,40]$ $\mu$m for
Model C1, $Z_1 = [60,65]$ $\mu$m for Model C2, and $Z_2 = [55,60]$
$\mu$m for Model C3. At the exterior boundary $\partial \Omega$, we apply the
boundary condition
\begin{equation}
    \label{eq:bcs:neumann}
    \Jvec_e^k \cdot \nvec_e = 0,
    \,  \text{at } \partial \Omega,
\end{equation}
describing that no ions can leave or enter the system.

In order to compare the KNP-EMI and the EMI framework, we set the
bulk conductivity $\sigma_r$ in the EMI model by~\eqref{eq:sigmar} with
initial values $[k]_r^0$ for the ion concentration $[k]_r$. Note that
$\sigma_r$ will generally change over time.

\subsubsection{Model D: Axon bundle with active Hodgkin-Huxley membrane model}

For model D, we consider a domain $\Omega = \Omega_i^1 \cup \dots \cup
\Omega_i^9 \cup \Omega_e = [0, 400] \times [0, 1.4] \times [0,1.4]$
$\mu$m, where 9 cuboidal cells of size $390 \times 0.2 \times 0.2$
$\mu$m are placed uniformly throughout $\Omega$
(cf.~Figure~\ref{fig:methods}). The distance between the cells is
$0.1$ $\mu$m. The domain is meshed as in Section~\ref{sec:modelA} with
$\Delta y = \Delta z = 0.5$ $\mu$m and $\Delta x = 0.625$ $\mu$m. The
ion channel currents are modelled using the Hodgkin-Huxley model as
described in Section~\ref{sec:mem}. An action potential is induced
every 20 ms throughout the simulations by applying the synaptic input
current model~\eqref{eq:synapse} with $g_{\text{syn}} = 4$
mS/cm\textsuperscript{2}, $\alpha = 2$ ms, and $t_0 = 0, 20, 40$
ms. We ran two sets of simulations: (1) stimulating, i.e.~applying the
synaptic current to the membrane of, the middle axon only (axon A in
Figure~\ref{fig:methods}), and (2) stimulating the 8 axons around axon
A (axons B,C in Figure~\ref{fig:methods}). At the exterior boundary we apply the boundary
conditions~\eqref{eq:bcs:dirichlet}.

\section{Results}
\label{sec:results}

We here present results from numerical experiments using the KNP-EMI
framework and the numerical method presented above. We start by
assessing the accuracy of the numerical method (model A and B). Next,
we compare the KNP-EMI and EMI frameworks in idealized 2D axons (model
C), before we finally investigate ephaptic coupling in unmyelinated
axons bundles (model D).

\subsection{Numerical verification and accuracy}

To evaluate the numerical accuracy and convergence of the proposed
numerical approach, we consider two sets of experiments.  First, we
examine the convergence of the model under mesh refinement by visual
inspection. Second, we perform a formal convergence analysis for a
smooth test case with manufactured solution.

\subsubsection{Qualitative inspection of convergence under mesh refinement}
\begin{figure}
    \centering
        \def\svgwidth{\columnwidth}
        \includegraphics[width=1.0\textwidth]{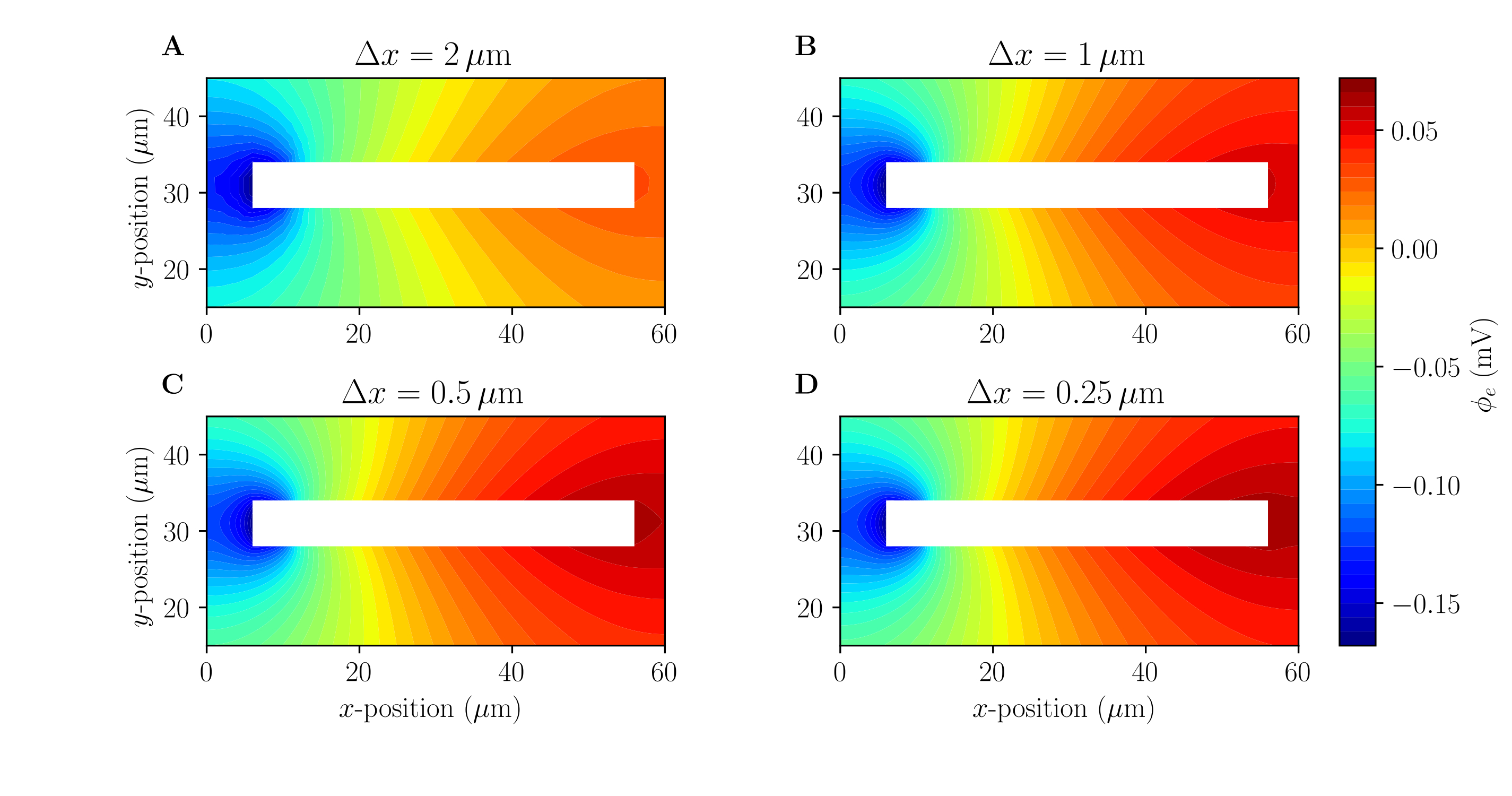}
        \def\svgwidth{\columnwidth}
        \includegraphics[width=1.0\textwidth]{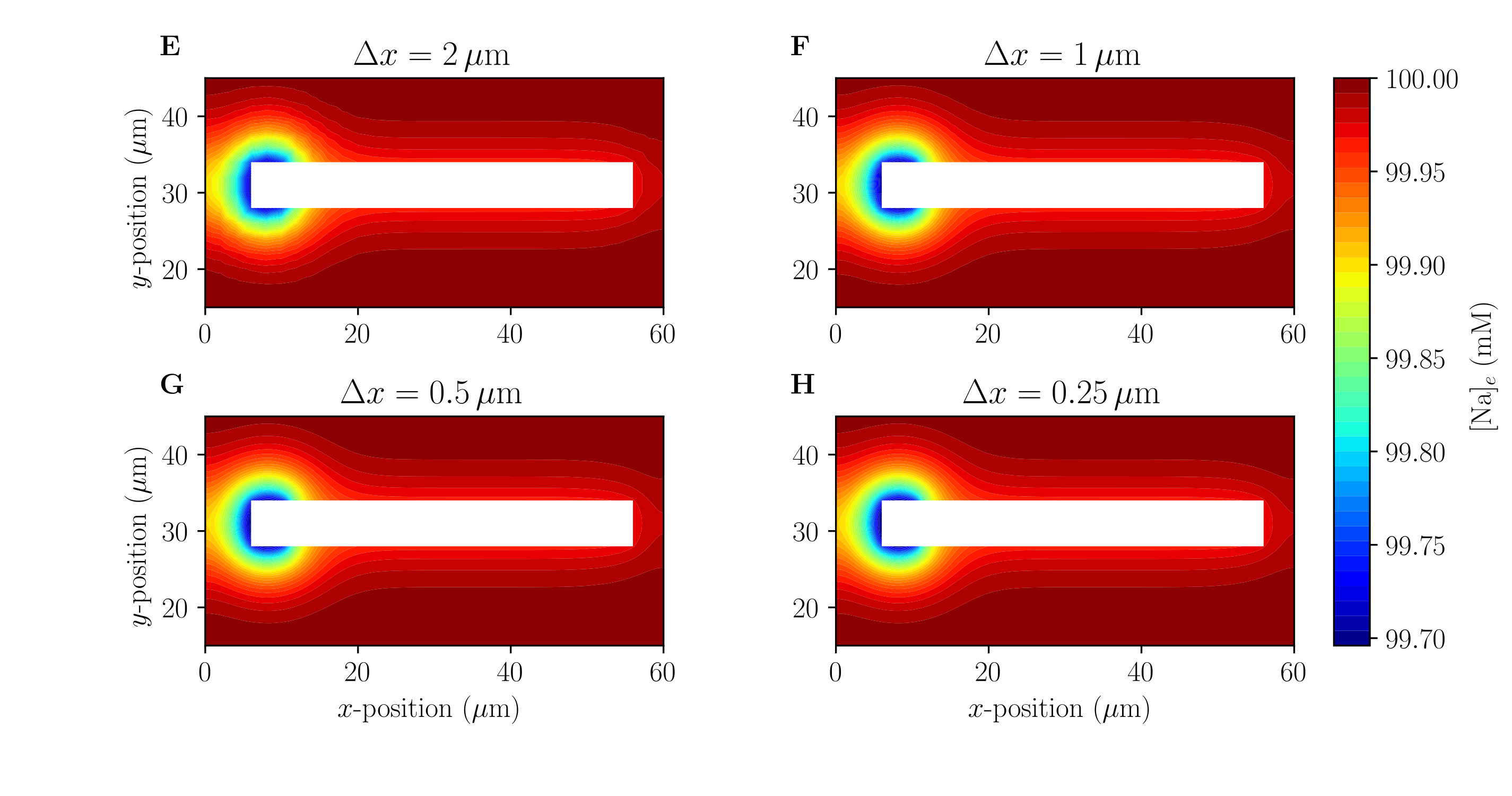}
    \caption{Model A: Comparison (under mesh refinement) of extracellular
    potential (\textbf{A}, \textbf{B}, \textbf{C}, \textbf{D}) and
    extracellular sodium concentration (\textbf{E}, \textbf{F},
    \textbf{G}, \textbf{H}) in the surroundings of a single simplified axon at
    $t = 10$ ms.}
\label{fig:refinement}
\end{figure}

The extracellular potential and sodium (\Naion)
concentration of Model A for four different mesh resolutions are shown at $t =
10$ ms in Figure~\ref{fig:refinement}. The
system quickly (\textcolor{red}{after 3 ms}) reaches a semi-steady state
where the membrane potential does not change notably over time, but there is a
slow exchange of sodium (\Naion) and potassium (\Kion) ions due to the leak
currents. The extracellular sodium concentration do not appear to change
visibly under mesh refinement, and the extracellular potential seems to reach a
converged state for the finest mesh resolution.

\subsubsection{Convergence rates of numerical solutions}
Using Model B, we analyzed the rates of convergence for the approximations of
all solution variables under refinement in space and time. Based on properties
of the approximation spaces and the time discretization, the optimal
theoretical rate of convergence is $1$ in the $H^1$-norm and $2$ in the
$L^2$-norm and the broken $L^2$-norm. Our numerical findings (Table
\ref{tab:convergence}) are in agreement with the theoretically optimal rates.
We observe second order convergence in the $L^2$-norm for the approximation of
the extracellular and intracellular concentrations and potentials, and first
order convergence in the $H^1$-norm. For the transmembrane ionic current $I_M$,
we observe a convergence rate of $1.5$ in the broken $L^2$-norm. The loss of
convergence of $\sim 0.5$ for $I_M$ is likely due to a lack of smoothness of
the interface in the test domain.
\begin{table}
    \begin{center}
        \begin{tabular}{c c c c c}
            \hline
            $n$
            & $\norm{[\Na]_i - [\Na]_{i, h}}_{0}$
            & $\norm{[\Na]_e - [\Na]_{e, h}}_{0}$
            & $\norm{[\Na]_i - [\Na]_{i, h}}_{1}$
            & $\norm{[\Na]_e - [\Na]_{e, h}}_{1}$ \\ \hline
            8 & 9.01E-03(---) & 3.12E-02(---) & 2.54E-01(---) & 8.80E-01(---) \\16 & 2.33E-03(1.95) & 8.08E-03(1.95) & 1.30E-01(0.97) & 4.50E-01(0.97) \\32 & 5.88E-04(1.99) & 2.04E-03(1.99) & 6.53E-02(0.99) & 2.26E-01(0.99) \\64 & 1.47E-04(2.00) & 5.10E-04(2.00) & 3.27E-02(1.00) & 1.13E-01(1.00) \\128 & 3.69E-05(2.00) & 1.28E-04(2.00) & 1.64E-02(1.00) & 5.67E-02(1.00) \\256 & 9.22E-06(2.00) & 3.21E-05(1.99) & 8.18E-03(1.00) & 2.86E-02(0.99) \\
            \hline \\ \hline 
            $n$
            & $\norm{[\K]_i - [\K]_{i, h}}_{0}$
            & $\norm{[\K]_e - [\K]_{e, h}}_{0}$
            & $\norm{[\K]_i - [\K]_{i, h}}_{1}$
            & $\norm{[\K]_e - [\K]_{e, h}}_{1}$ \\ \hline
            8 & 9.01E-03(---) & 1.04E-02(---) & 2.54E-01(---) & 2.93E-01(---) \\16 & 2.33E-03(1.95) & 2.69E-03(1.95) & 1.30E-01(0.97) & 1.50E-01(0.97) \\32 & 5.88E-04(1.99) & 6.79E-04(1.99) & 6.53E-02(0.99) & 7.54E-02(0.99) \\64 & 1.47E-04(2.00) & 1.70E-04(2.00) & 3.27E-02(1.00) & 3.78E-02(1.00) \\128 & 3.69E-05(2.00) & 4.25E-05(2.00) & 1.64E-02(1.00) & 1.89E-02(1.00) \\256 & 9.22E-06(2.00) & 1.20E-05(1.82) & 8.18E-03(1.00) & 1.02E-02(0.89) \\
            \hline \\ \hline 
            $n$
            & $\norm{[\Cl]_i - [\Cl]_{i, h}}_{0}$
            & $\norm{[\Cl]_e - [\Cl]_{e, h}}_{0}$
            & $\norm{[\Cl]_i - [\Cl]_{i, h}}_{1}$
            & $\norm{[\Cl]_e - [\Cl]_{e, h}}_{1}$ \\ \hline 
            8 & 1.80E-02(---) & 4.16E-02(---) & 5.08E-01(---) & 1.17E+00(---) \\16 & 4.67E-03(1.95) & 1.08E-02(1.95) & 2.60E-01(0.97) & 6.00E-01(0.97) \\32 & 1.18E-03(1.99) & 2.72E-03(1.99) & 1.31E-01(0.99) & 3.02E-01(0.99) \\64 & 2.95E-04(2.00) & 6.82E-04(2.00) & 6.54E-02(1.00) & 1.51E-01(1.00) \\128 & 7.38E-05(2.00) & 1.71E-04(1.99) & 3.27E-02(1.00) & 7.56E-02(1.00) \\256 & 1.84E-05(2.00) & 4.48E-05(1.93) & 1.64E-02(1.00) & 3.85E-02(0.97) \\
            \hline \\ \hline 
            $n$
            & $\norm{\phi_i - \phi_{i, h}}_{0}$
            & $\norm{\phi_e - \phi_{e, h}}_{0}$
            & $\norm{\phi_i - \phi_{i, h}}_{1}$
            & $\norm{\phi_e - \phi_{e, h}}_{1}$ \\ \hline
            8 & 5.83E-02(---) & 1.43E-01(---) & 1.69E+00(---) & 1.43E+00(---) \\16 & 1.61E-02(1.86) & 3.81E-02(1.91) & 8.66E-01(0.96) & 7.43E-01(0.94) \\32 & 4.13E-03(1.96) & 9.67E-03(1.98) & 4.35E-01(0.99) & 3.76E-01(0.98) \\64 & 1.04E-03(1.99) & 2.43E-03(2.00) & 2.18E-01(1.00) & 1.89E-01(1.00) \\128 & 2.54E-04(2.03) & 6.03E-04(2.01) & 1.09E-01(1.00) & 9.44E-02(1.00) \\256 & 5.90E-05(2.11) & 1.44E-04(2.07) & 5.45E-02(1.00) & 4.74E-02(0.99) \\
            \hline \\ \hline 
            $n$
            & $\norm{I_M - I_{M,h}}_{0, \Gamma}$ \\ \hline 
            8 & 7.03E+00(---) \\16 & 2.54E+00(1.47) \\32 & 8.93E-01(1.51) \\64 & 3.14E-01(1.51) \\128 & 1.11E-01(1.50) \\256 & 3.94E-02(1.49) \\ 
            \hline
        \end{tabular}
    \end{center}
    \caption{Approximation errors (with convergence rates in parenthesis) for
    the extracellular and intracellular concentrations and potentials, and
    transmembrane current, under simultaneous refinement in time and space.
    Approximation errors are measured at $t = \frac{2}{64} \times 10^{-7}$,
    i.e.~e.g.~$\|[\Na]_r - [\Na]_{r, h}\|_0 = \|[\Na]_r(t) - [\Na]_{r,
    h}(t)\|_0$ and similarly for all reported values.} 
    \label{tab:convergence}
\end{table}

\subsection{Comparison of the KNP-EMI and EMI frameworks in idealized axons}
The KNP-EMI framework extends the EMI framework by explicitly
modelling the ionic concentrations and incorporating ionic
electrodiffusion. A key question is when and to what extent the
solutions from the two (KNP-EMI and EMI) frameworks differ. To compare
the two frameworks, we consider three models (Model C1, C2 and C3) and
compare the corresponding solution of the KNP-EMI equations (the
KNP-EMI solution) with the solution of the EMI equations (the EMI
solution).

\begin{figure}
  \centering 
    \def\svgwidth{\columnwidth}
    \includegraphics[width=1.0\textwidth]{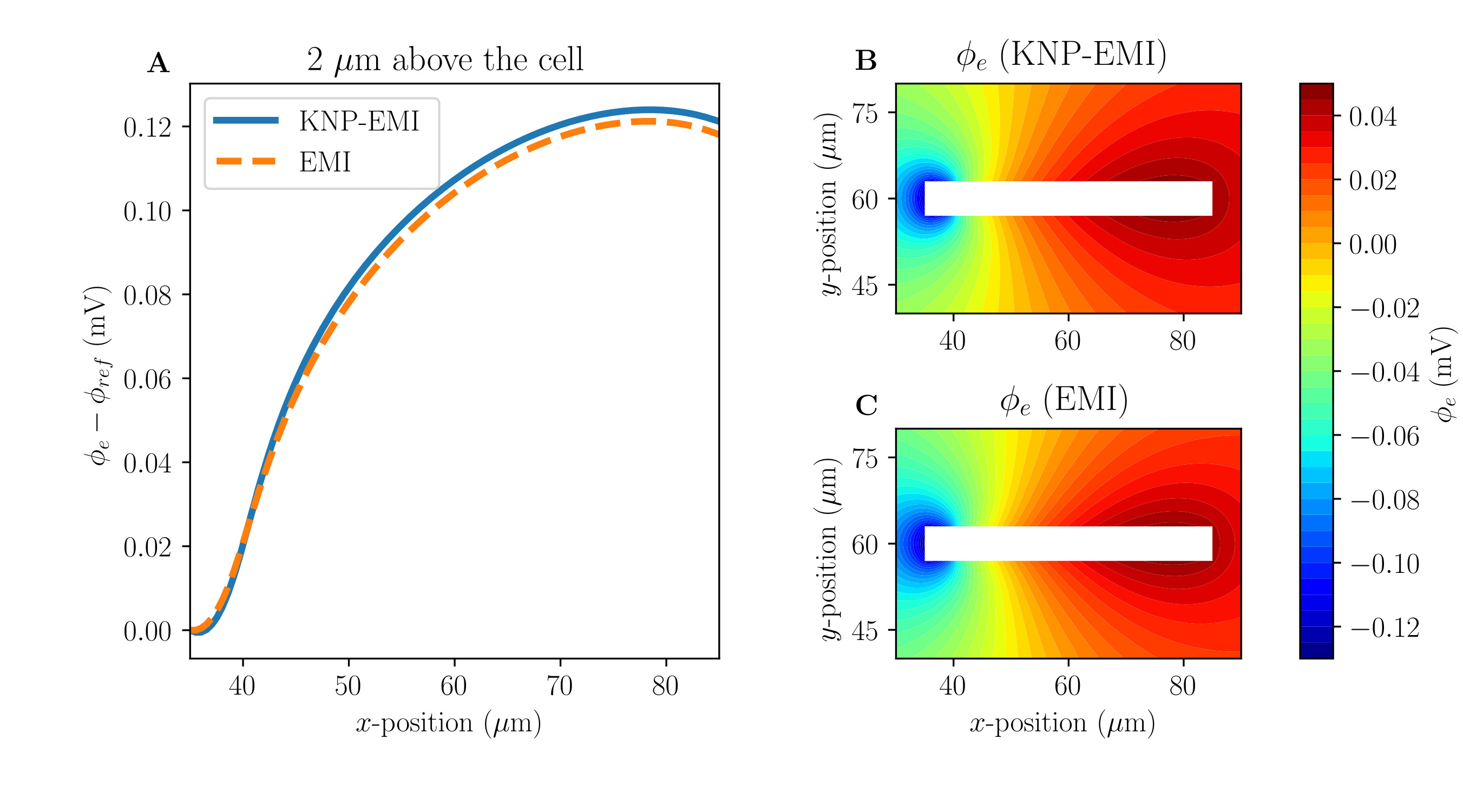}
    \caption{A comparison of the KNP-EMI and the EMI frameworks using model C1 at
    $t=10$ ms. Normalized (to have the same value at $x=35 \mu$m) extracellular potentials
    measured $2 \mu$m above the cell
    (\textbf{A}). Extracellular potentials from the KNP-EMI (\textbf{B}) and
    the EMI framework (\textbf{C}) surrounding the cell.}
  \label{fig:one_neuron}
\end{figure}

We first consider the extracellular potential resulting from
stimulating a single axon (Model C1) using the KNP-EMI and EMI
frameworks (Figure \ref{fig:one_neuron}). We observe that the KNP-EMI
and EMI solutions are qualitatively very similar: an extracellular
potential difference of approximately $0.12$ mV along the length of
the axon develops in both (Figure \ref{fig:one_neuron}\,A, B, C).

\begin{figure}
    \centering
    \def\svgwidth{\columnwidth}
    \includegraphics[width=1.0\textwidth]{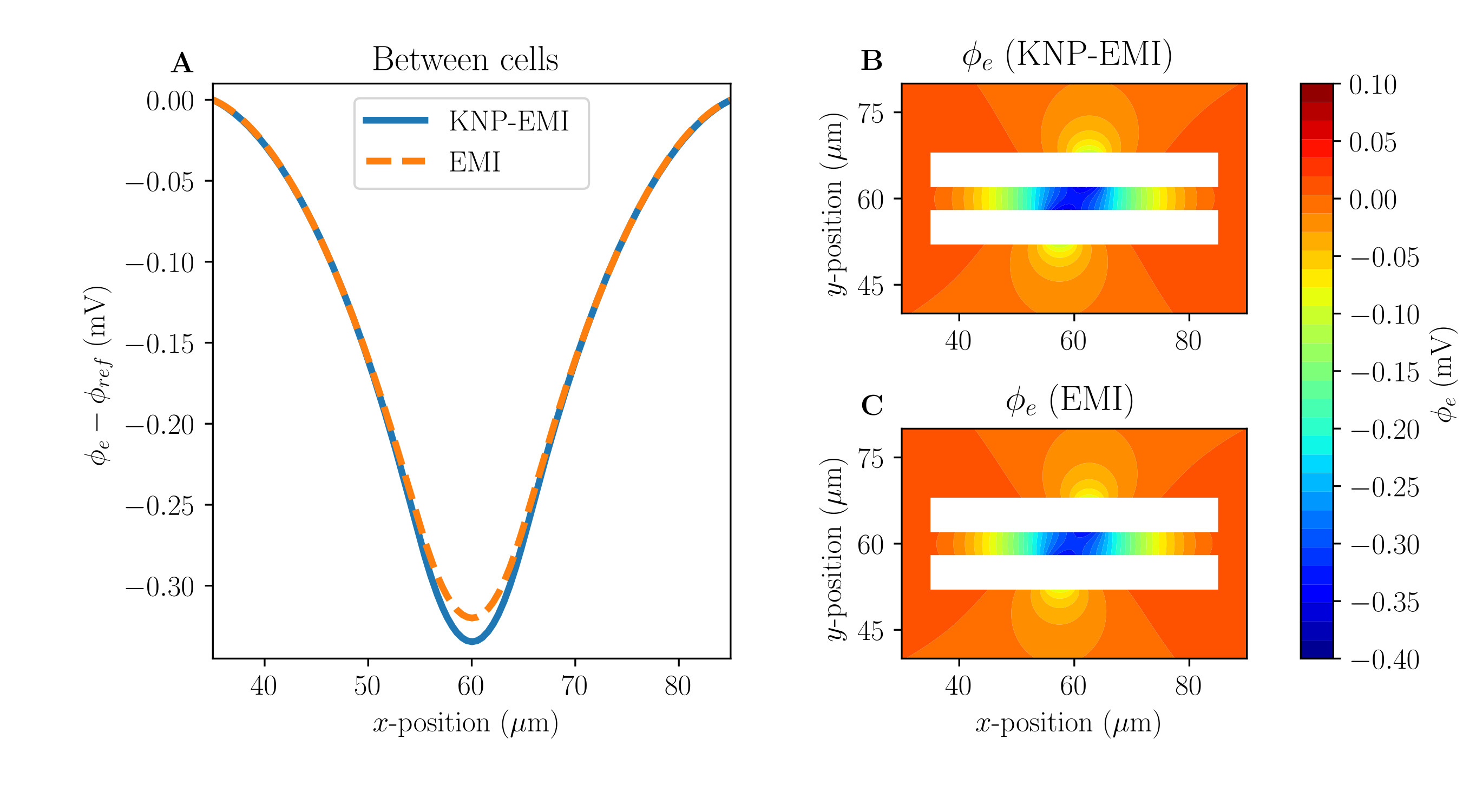}
    \def\svgwidth{\columnwidth}
    \includegraphics[width=1.0\textwidth]{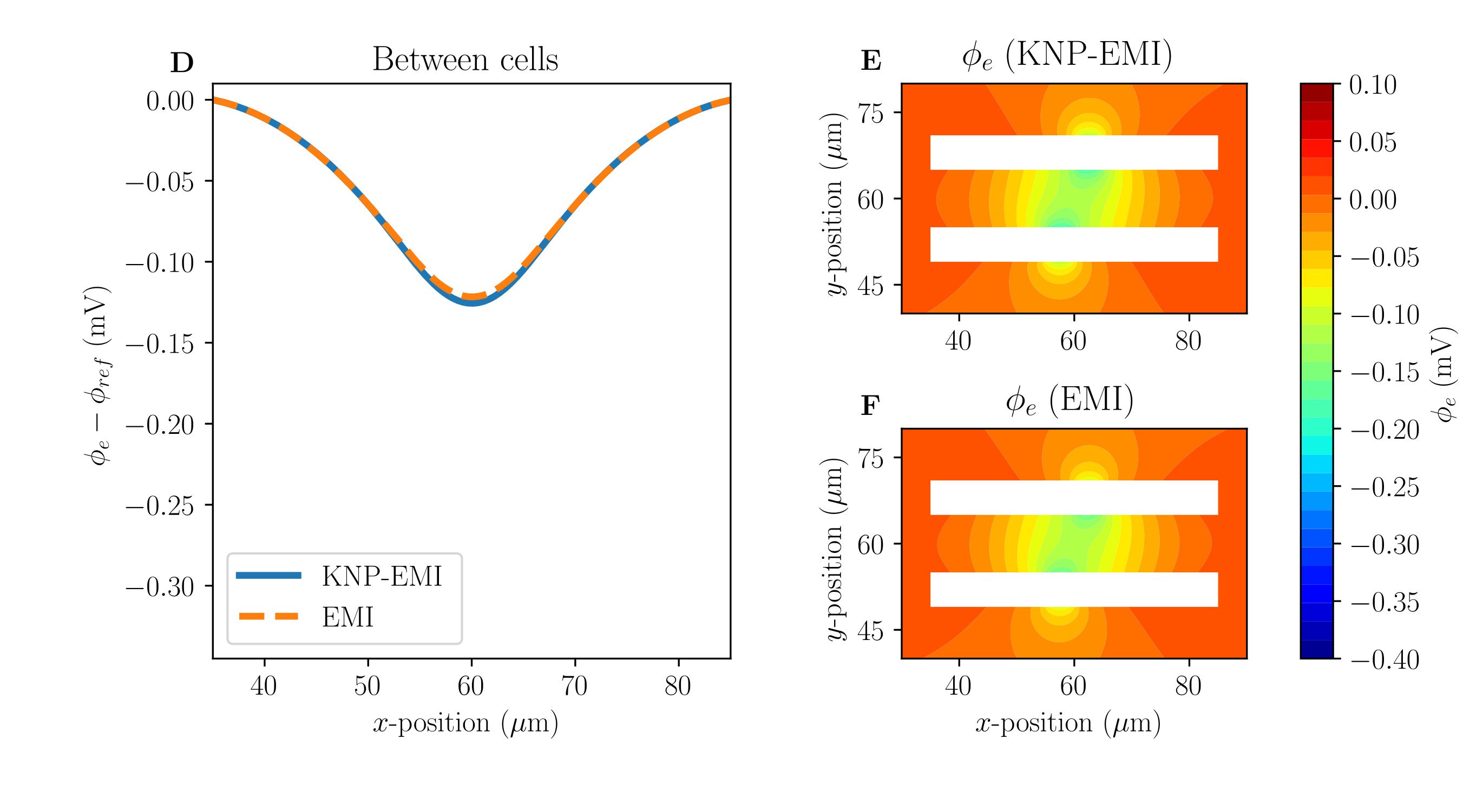}
    \caption{A comparison of the KNP-EMI and the EMI frameworks using
      Model C2 and C3 at $t=10$ ms. The normalized (to have the same
      value at $x = 35 \mu$m) extracellular potentials from
      Model C2 (\textbf{A}) and C3 (\textbf{D}) on the midline between
      the neurons ($y = 60 \mu$m). The extracellular potentials
      surrounding the cells as calculated by KNP-EMI
      (\textbf{B}) and EMI (\textbf{C}) using Model C2, and by KNP-EMI
      (\textbf{E}) and EMI (\textbf{F}) using Model C3.}
    \label{fig:two_neurons_and_wide}
\end{figure}

Next, we compare the extracellular potentials resulting from stimulating two
neighboring axons (Model C2 and C3) using the KNP-EMI and the EMI frameworks
(Figure \ref{fig:two_neurons_and_wide}). The two models differ by the distance
between the axons. For Model C2, we again observe that the KNP-EMI and EMI
solutions match closely, but differ locally. The maximal difference between the
extracellular potential solutions is \textcolor{red}{$0.02$} mV (Figure
\ref{fig:two_neurons_and_wide}\,A--C). For Model C3, we observe the analogous
behaviour, but note that the extracellular field is weaker than for
Model C2 (Figure \ref{fig:two_neurons_and_wide}\,D--F).

\subsection{Ephaptic coupling in unmyelinated axon bundles}
We now turn to explore the effect of ephaptic coupling in an idealized
axon bundle with 9 axons using the KNP-EMI framework. 
We consider two sets of simulations using Model D: (1) stimulating,
i.e.~applying the synaptic current to the membrane of, the middle axon
only (axon A, Figure~\ref{fig:methods}), and (2) stimulating the 8
axons around axon A (axons B,C, Figure~\ref{fig:methods}).

\subsubsection{Electrodiffusion effects in unmyelinated axon bundles}
\begin{figure}
    \centering
    \def\svgwidth{0.80\columnwidth}
    \includegraphics[width=0.8\textwidth]{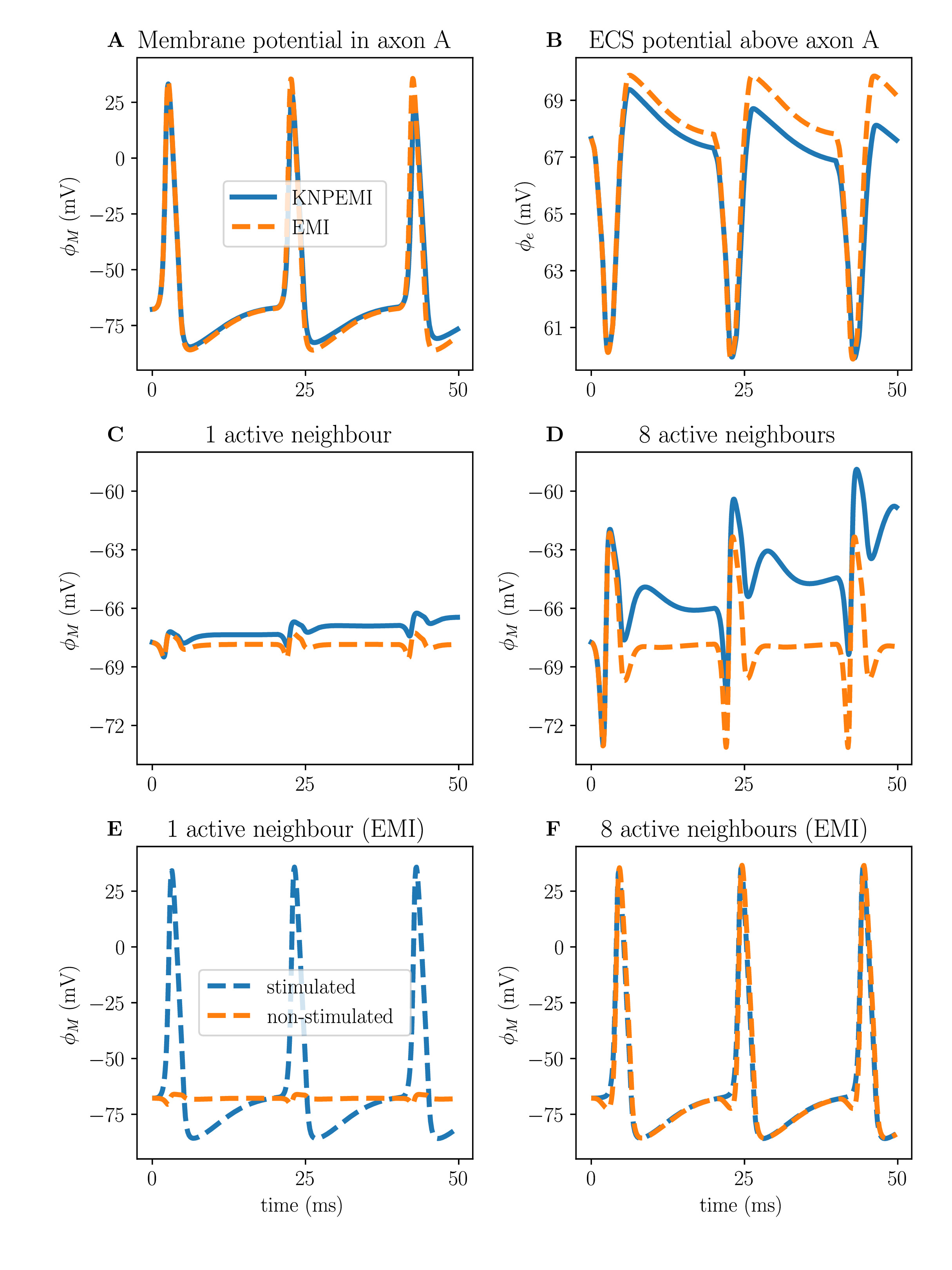}
    \caption{Effects of ephaptic coupling in a bundle of axons at $x=200$ $\mu$m.
        The membrane potential ($\phi_M$) of axon A (\textbf{A}), and the
        extracellular potential ($\phi_e$) measured 0.05 $\mu$m away from the
        membrane of axon A (\textbf{B}) during stimuli of axon A only.  Ephaptic
        coupling measured in axon B when only axon A is stimulated (\textbf{C}),
        and measured in axon A when all peripheral axons (B--C) are stimulated
        (\textbf{D}).  Setting $\sigma_i = 1.0$ $\mu$S/$\mu$m and $\sigma_e = 0.1$
        $\mu$S/$\mu$m in the EMI framework increases the ephaptic coupling to the point
        where simultaneous action potentials in all 8 surrounding axons will induce
        an action potential in the central axon (\textbf{F}). However, only
        stimulating the middle axon (A) will not induce action potentials in the
        peripheral axons (\textbf{E}).}
    \label{fig:ephaptic}
\end{figure}

To investigate ephaptic coupling, we first apply a synaptic current to
stimulate the cell membrane of the middle axon of the axon bundle
(axon A, Figure~\ref{fig:methods}, Model D). The synaptic current
induces a series of action potentials in axon A and also induces
substantial changes in the surrounding extracellular potential
(Figure~\ref{fig:ephaptic}\,A, B). The extracellular potential
fluctuations further spread to axon B. However, the ephaptic effect on
the membrane potential in axon B is relatively small (1--2 mV), and is
not sufficient to reach the threshold for inducing an action potential
(Figure~\ref{fig:ephaptic}\,C).

The ephaptic effect is stronger if we simultaneously stimulate the
cell membranes of all 8 peripheral axons (axons B--C). Again, we
observe a series of action potentials in the 8 stimulated
axons. Moreover, the combined ephaptic currents have a pronounced
excitatory effect on axon A, but again fail to induce an action
potential there (Figure~\ref{fig:ephaptic}\,D).

\begin{figure}
    \centering 
    \def\svgwidth{0.80\columnwidth}
    \includegraphics[width=0.8\textwidth]{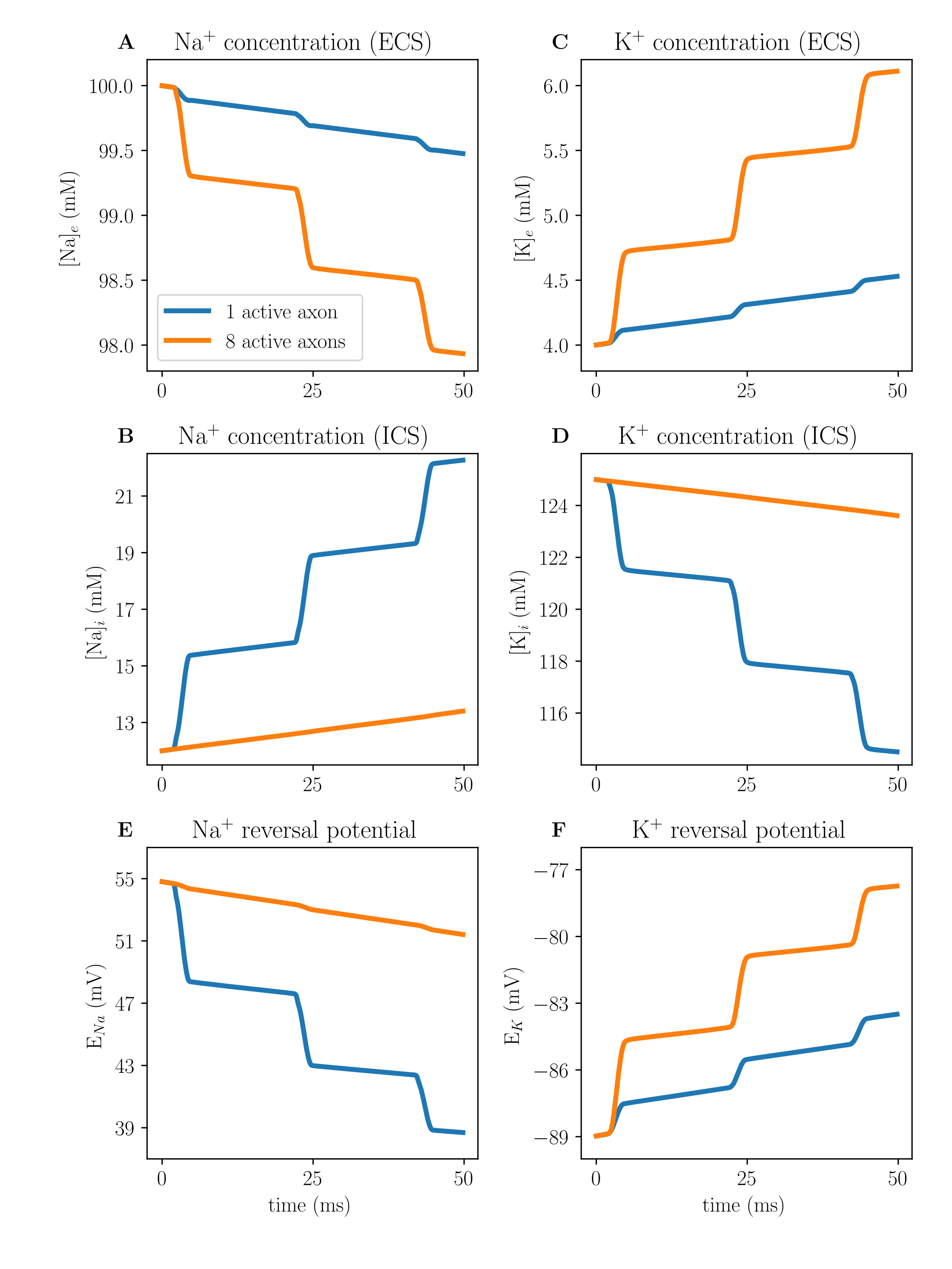}
    \caption{Ion concentration dynamics in an axon bundle measured at the
        middle axon (A) at $x=200$ $\mu$m using the KNP-EMI framework, both when
        middle axon (A) is stimulated, and when all peripheral axons (B--C) are
        stimulated.  Extracellular sodium (\textbf{A}) and extracellular potassium
        (\textbf{B}) concentrations evaluated 0.05 $\mu$m away from axon A.
        Intracellular sodium (\textbf{C}) and intracellular potassium (\textbf{D})
        concentrations evaluated at the center of axon A.  Reversal potentials for
        sodium (\textbf{E}) and potassium (\textbf{F}) at the membrane of axon
        A.}
        \label{fig:concentrations}
\end{figure}

The difference between the EMI and KNP-EMI simulations are due to the
time evolution of the intracellular and extracellular ion
concentrations, accounted for by the KNP-EMI model but not by the EMI
model. For each action potential fired, the Nernst potential will
change due to alterations in the ionic concentrations using the
KNP-EMI framework (Figure~\ref{fig:concentrations}), whereas in the
EMI framework the Nernst potential is constant.

Our predictions differ from those made in a similar study by Bokil et
al~\cite{bokil2001ephaptic}, who found that a single active neighbor
can induce action potentials in all nearby axons. We hypothesize that
the main explanation for these differences is that the bulk
conductivities differ between the two studies. Here, in the KNP-EMI
framework, the bulk conductivities are functions of the ion concentrations
(cf.~\eqref{eq:sigmar}). Using realistic values for the intra- and
extracellular ion concentrations, we obtained bulk conductivities values of
$\sigma_i \approx 2.01 \mu$S/$\mu$ m and $\sigma_e \approx
1.31 \mu$S/$\mu$m. In contrast, Bokil et al.~set the bulk conductivities as
free parameters, with $\sigma_i = 1 \, \mu$S/$\mu$m and $\sigma_e =
0.1 \, \mu$S/$\mu$m as the corresponding effective bulk conductivities in the
EMI model. Tveito et al.~\cite{tveito2017evaluation} found that the
ephaptic current was inversely proportional to $\sigma_e$, which
suggests that the ephaptic current was more than 7 times stronger in
Bokil et al.~\cite{bokil2001ephaptic} than here.

In light of this, we repeated the simulations of the EMI model using
the lower effective bulk conductivity values ($\sigma_i = 1 \, \mu$S/$\mu$m
and $\sigma_e = 0.1 \, \mu$S/$\mu$m). In this case, simultaneous
stimulation of the 8 peripheral axons (B--C) induced an action
potential in axon A (Figure~\ref{fig:ephaptic}\,F). Stimulation of
axon A alone did not induce an action potential in the 8 peripheral
axons (Figure~\ref{fig:ephaptic}\,E).

\section{Discussion}

We have presented a finite element-based numerical method for a
revised mathematical model of ionic electrodiffusion with explicit
geometrical representation of the extracellular space, the
intracellular space and the cell membrane. Our numerical scheme is
based on the mortar finite element method and is capable of
efficiently handling complex geometries in one, two or three spatial
dimensions. Our numerical investigations demonstrate that the scheme
is accurate and yields optimal convergence rates in the relevant
norms.

Further, we compared the KNP-EMI framework and the EMI framework by
computationally studying (i) extracellular fields surrounding passive
idealized axons, and (ii) membrane potentials in a bundle of
unmyelinated axons under Hodgkin-Huxley membrane mechanisms. The
potentials predicted by the two frameworks are essentially identical
during the first period ($\sim$$5$ ms) of the simulations, but the
predictions later differ due to changes in ion concentrations (only
accounted for by the KNP-EMI framework). We note that the strongest
ephaptic coupling is due to changes in the Nernst potentials (ionic
ephaptic coupling), and not via extracellular potentials (electric
ephaptic coupling).

The predictions of ephaptic coupling made in this study differs from those made
by Bokil et al~\cite{bokil2001ephaptic} using cable theory. This discrepancy is
likely due to differences in the extracellular bulk conductivities. Indeed, an
important difference between geometrically explicit frameworks (e.g.~PNP, EMI
and KNP-EMI) and homogenized frameworks (e.g.~cable theory) is the
interpretation of the bulk conductivities $\sigma_i$ and $\sigma_e$. In
homogenized frameworks based on volume-conductor theory, the bulk conductivity
$\sigma$ is interpreted as the tissue average, i.e.~the effective bulk
conductivity for currents propagating over distances in brain
tissue~\cite{Holt1999, Pettersen2008, Reimann2013}. Importantly, this
tissue-averaged bulk conductivity is smaller than the actual conductivity of
the extracellular solution, largely due to the fact that the extracellular
space only constitutes about 20\% of the total tissue volume. On the other
hand, in the KNP-EMI framework, the bulk conductivities are defined in terms of
the local ion concentrations and will thus vary consistently across the domain.

The KNP-EMI framework is easily comparable to the PNP
framework~\cite{lopreore2008computational, pods2013electrodiffusion,
holcman2015new, cartailler2017electrostatics, cartailler2017analysis}
as both frameworks can account for the explicit morphology of neural
tissue~\cite{biess2007diffusion,noguchi2005spine}. The PNP scheme uses
the PNP formalism for all tissue components, which requires the
resolution of the charge accumulation in the Debye layers using a fine
spatiotemporal resolution. As a result, PNP schemes are well suited to
make predictions on fine spatial scales close to cellular membranes,
but longer simulations on larger domains are not computationally
feasible. In contrast, in the KNP-EMI framework we circumvent the need
for explicit modelling of charge accumulation near the membrane by
assuming electroneutrality.  This results in a more numerically stable
framework for coarser time and space resolutions, allowing for longer
simulations on larger domains. The differences between the PNP
framework and electroneutral frameworks, such as KNP, have been
discussed extensively in previous works~\cite{mori2009numerical,
pods2017comparison, solbra2018kirchoff}.

An example of a phenomenon where large ion concentration changes in
brain tissue build up over time, is (cortical) spreading
depression. During spreading depression, the extracellular
K\textsuperscript{+} concentration can change from a basal level of
3-5 mM to peak values at tens of mM over a period of several
minutes~\cite{Somjen2001}. As such, we advocate that the KNP-EMI model
would be suitable for studying cellular level aspects of spreading
depression computationally. However, for simulations of longer
duration ($>50$ ms), the membrane mechanism model should be chosen
carefully. The Hodgkin-Huxley formalism used in this paper to describe
the membrane mechanisms does not account for the effect of ion pumps
and co-transporters, which generally will strive to restore
concentrations to baseline. As a consequence, the concentration
changes in Figure~\ref{fig:refinement}--\ref{fig:ephaptic} are likely
overestimates of what could be expected to occur at such short time
scales. Adding ion pumps and co-transporters to the membrane model
would be relatively
straightforward~\cite{hubel2014bistable,hubel2014dynamics}.

In conclusion, the KNP-EMI framework presented in this paper allows
for detailed computational studies of the interplay between ion
movement, membrane mechanisms and electrical potential in healthy
neural tissue and under pathological conditions. The computational
expense of KNP-EMI simulations compared to e.g.~homogenized models
calls for further research into efficient and scalable solution
methods.

\section*{Conflict of Interest Statement}
The authors declare that the research was conducted in the absence of any commercial or financial relationships that could be construed as a potential conflict of interest.

\section*{Author Contributions}

AJE, AS, GTE, GH, and MER designed the study. AJE, AS, GTE, GH and MER
derived the mathematical model and numerical method. AJE and AS
developed the software implementation and performed the numerical
experiments. AJE, AS, GTE, GH and MER wrote the manuscript. 

\section*{Funding}
This project has received funding from the European Research Council
(ERC) under the European Union's Horizon 2020 research and innovation
programme under grant agreement 714892 (Waterscales), and from 
the Research Council of Norway (BIOTEK2021 Digital Life project
`DigiBrain', project 248828).

\section*{Acknowledgments}
We thank Klas Pettersen (NORA) and Cécile Daversin-Catty (Simula
Research Laboratory) for useful discussions.

\section*{Data Availability Statement}
Our implementation of the finite element scheme presented in this paper is
publicly available~\cite{CodeZenodoDoi}.

\newpage
\FloatBarrier

\bibliographystyle{frontiersinHLTHFPHY}
\bibliography{manuscript}

\end{document}